\documentclass[fleqn,usenatbib]{mnras}

\usepackage{newtxtext,newtxmath}

\usepackage[T1]{fontenc}

\DeclareRobustCommand{\VAN}[3]{#2}
\let\VANthebibliography\thebibliography
\def\thebibliography{\DeclareRobustCommand{\VAN}[3]{##3}\VANthebibliography}

\usepackage{graphicx}	
\usepackage{amsmath}	
\usepackage[dvipsnames]{xcolor}

\usepackage{siunitx}
\usepackage{array}

\usepackage{url}

\usepackage{ulem}  

\graphicspath{{Images/}}

\title[PRG-galaxy catalog]{ Peculiar galaxies I: A Catalog of Polar-Ring Galaxies from the TNG50 Simulation}

\author[J. G. López-Castillo et al.]{
Josué G. López-Castillo,$^{1}$\thanks{E-mail: josue.gerardo@inaoep.mx} 
Manuel Zamora-Avilés,$^{1}$ 
Gilberto C. Gómez,$^{2}$ 
Ivânio Puerari,$^{1}$  
\newauthor
and Divakara Mayya$^{1}$ 
\\
$^{1}$Instituto Nacional de Astrofísica, Óptica y Electrónica, Luis E. Erro 1, 72840 Tonantzintla, Puebla, México\\
$^{2}$Instituto de Radioastronomía y Astrofísica, Universidad Nacional Autónoma de México, Apdo. postal 3-72, Morelia Mich. 58089, México\\
}

\date{Accepted XXX. Received YYY; in original form ZZZ}

\pubyear{2015}

\begin{document}
\label{firstpage}
\pagerange{\pageref{firstpage}--\pageref{lastpage}}
\maketitle

\begin{abstract}
    The hydrodynamic cosmological simulation, TNG50, is employed to conduct an analysis of multi-spin galaxies that exhibit ringed structures composed of gas and stars that orbit nearly perpendicular around a host galaxy, known as polar ring galaxies (PRG). To ensure a robust sample, we select subhalos based on the angle subtended by the angular momentum profiles, as well as on a visual inspection. The analysis is focused on galaxies with stellar masses greater than 10$^{9}$ M$_\odot$. In addition, a dynamic decomposition is employed to separate the stellar and gaseous ring from the host galaxy. This results in a sample of 32 subhalos with PRGs. This sample exhibits properties similar to those observed. These include colours typical of early-type galaxies (ETGs) or those transitioning toward blue systems. Most host galaxies are classified as ETGs, with 37.5\% exhibiting a disk-dominated morphology. The mean bulge-to-total (B/T) ratio is 0.64. Rings have average radii that are 2.36 and 3.41 times larger than their effective radii for the stellar and gaseous components, respectively, with star formation occurring predominantly within the rings. In contrast with observations, rings in the simulation tend to be less massive and slightly less perpendicular. The obtained sample displays a variety of host galaxy morphologies, including wide and narrow rings, providing a robust framework for studying the varied structural characteristics of PRG variants.
\end{abstract}




\section{Introduction}
The study of peculiar galaxies enables us to test our models of the universe by examining objects that form and evolve under extreme environmental conditions.
{Polar Ring Galaxies (PRGs) are peculiar systems characterised by a central host galaxy encircled by an outer ring of stars, gas, and dust that orbits nearly perpendicular to the midplane of the host \citep{1990_Whitmore}. The host galaxy in these objects is predominantly associated with early-type galaxies (ETGs) \citep{1995_Arnaboldi,2002I_Iodice,Reshetnikov_2015}.
The wide range of structures exhibited by PRGs makes them particularly valuable for studies of galaxy evolution. This diversity includes thin and compact structures, as well as very extended rings \citep{1997_Reshetnikov, 2012_Finkelman, 2016_Ordenes-Briceno}. Some of these structures are even disk-like, showing features such as spiral arms \citep[e.g., NGC4650A;][]{1997_Reshetnikov,2002c_Iodice, 2002_Gallagher, 2006_Iodice}. A significant fraction of polar ring structures appear bluer and fainter than their host galaxies, making them detectable only in deep galaxy surveys \citep{Reshetnikov_2015, 2024_Mosenkov}. A possible correlation between PRGs and nuclear activity has also been suggested. While an excess of Seyfert and LINER galaxies among PRGs has been reported, other observational studies find no significant correlation compared to ``normal'' ETGs \citep{2001_Reshetnikov,2012_Finkelman, 2020_Smirnov}.

In this context, PRGs have been used to better understand the distribution of dark matter in galaxies \citep{2012_Snaith,2014_Khoperskov}, the formation and evolution of galaxies, and the impact on the dark matter halo \citep{2006_Maccio}, or the nature of the Tully-Fisher relation \citep{2003_Iodice}.}
{At the same time, the ring contains a substantial reservoir of gas comparable in mass to that of the host galaxy, and its colour indicates one or more recent episodes of star formation \citep{1997_Galletta,2002_Iodice,Reshetnikov_2015,2024_Mosenkov}. A connection between PRGs and active nuclei is also expected, although some observational results have not found a significant correlation compared to ``normal'' ETGs \citep{2001_Reshetnikov,2012_Finkelman}. Moreover, PRGs tend to reside in low-density environments, which helps preserve their rings by reducing disruptive interactions with neighbouring galaxies \citep{2012_Finkelman,2017_Savchenko, 2020_Smirnov}. Unfortunately, only a few dozen PRGs have been confirmed, largely due to their diffuse nature and orientation effects that complicate identification. These limitations, in turn, constrain the robustness of the derived results. \citet{1990_Whitmore} estimated that about 5\% of nearby lenticular galaxies have or have had polar structures, while \citet{2011_Reshetnikov} found a corresponding fraction of $\sim$0.4\% for nearby galaxies on the B band. \citet{2022_Smirnov} reported a lower fraction ($\sim$0.01\%) in the $r$ band, increasing with redshift. More recently, \citet{2024_Mosenkov} estimated a PRG fraction of $\sim$3\% in the $r$ band, considering projection effects.
}  

Several attempts have been made to understand the origin of these objects, with several scenarios proposed and tested using numerical simulations for comparison with observational data.
One of the proposed scenarios, known as the merge scenario, consists of a head-on collision of two disk galaxies, similar to the formation of collisional ring galaxies but with low velocities, where the perturber galaxy ends as the host galaxy. \cite{1997_Bekki,1998_Bekki} demonstrated that PRGs formed by the merge scenario result in stable systems. For reference, AM 2229-735 has been classified as a PRG in formation through minor mergers \citep{2020_Quiroga}. Another proposed scenario, the accretion scenario, consists of two galaxies undergoing a tidal interaction; one of them, the host galaxy, accretes gas from the donor galaxy that orbits almost perpendicular to the host galaxy. In this scenario, the interaction may or may not end in a merger \citep{1983_Schweizer,1997_Reshetnikov,2003_Bournaud}. A representative example of this formation scenario is NGC 3808B, which exhibits a tidal gas bridge being transferred from its companion NGC 3808A \citep{1996_Reshetnikov, 2016_Ordenes-Briceno}. Yet another model, the cosmological scenario, proposes that polar rings are formed by the accretion of cold gas from cosmological filaments onto the host galaxy. In this case, the accretion occurs perpendicular to the central galaxy \citep{2006_Maccio,2008_Brook,2012_Snaith}. An observational signature of PRGs formed under this scenario is the low metallicity in their rings, indicative of the late infall of metal-poor gas, as is expected in cold accretion processes. This has been observed in systems such as NGC4650A \citep{2010_Spavone} and others \citep[eg. UGC 7576, UGC 9796 or  IIZw71;][]{2011_Spavone, 2019_Egorov}.
There is no definite explanation for the origin of PRGs, as all of these scenarios likely occur in nature. Therefore, identifying observational features that can distinguish among the various formation mechanisms in individual galaxies would be important to assess the relative contributions of these different pathways.

Today, a variety of cosmological simulations have been performed, for example, EAGLE \citep{2015_Crain,2015_Schaye}, IllustrisTNG \citep{Pillepich_2019, Nelson_2019}, or Horizon-AGN \citep{2014_Dubois}. These simulations can reproduce observational constraints such as star formation rates, galaxy sizes, and dynamics \citep{2018_Nelso, Genel_2017, Pillepich_2019}. Thus, cosmological simulations have become a powerful tool for constraining the model of galaxy formation and evolution. 

Historically, galaxies have been classified according to their apparent projected morphology, allowing us to catalogue them, for example, according to the distribution of the bulge and the disk \citep{1926_Hubble,1968_Sersic}. However, the advantage of numerical simulations is that they provide us with information about the morphological and dynamical parameters of galaxies over time. In order to accurately study galaxies in cosmological simulations, it is necessary to use methods that split them into their constituent parts. 

Recent studies have focused on clarifying the origins and evolutionary processes of these objects. \citet{2023_Smirnov} conducted an analysis of PRGs selected through visual inspection of the TNG50 Infinite Gallery page\footnote{\url{https://www.tng-project.org/explore/gallery/rodriguezgomez19b/}} at redshift 0.05. Six galaxies with the characteristics of PRGs were selected for further analysis. In these galaxies, the polar structure is formed by interaction with a gas-rich companion or satellite. The formed rings evolve and change their inclination with respect to the disk. Additionally, \citet{2023_Smirnov} reported the relation of the formation of polar rings and temporary bursts of nuclear activity. Other types of annular structures have been identified in the EAGLE simulation by \citet{2018_Elegali}, who classified ring galaxies into two categories: O-type (resonance ring galaxies) and P-type (collisional ring galaxies). In contrast, as discussed in Section \ref{sec: Methods}, our selection scheme specifically targets misaligned ring structures around galaxies.

The aim of this paper is the identification of PRGs in the hydrodynamic cosmological simulation TNG50 \citep{Pillepich_2019, Nelson_2019}, which has the best resolution in the IllustrisTNG simulations in an extended cosmological box. This allows us to identify and study a consistent sample of PRGs and compare it with the observational counterpart. The simulations also allow for the analysis of galaxy dynamics and the decomposition of galaxies into their different components, both in terms of their stellar and gaseous structure. Thus, the formation and evolution of the rings can be characterised. This sample of PRGs can be useful for future work to constrain the origin of these peculiar galaxies as well as their observables. The structure of this paper is as follows. A brief description of the cosmological simulation used, and the method used to select the PRGs as well as the decomposition method used to detect the ring structures is given in \ref{sec: Methods}. Section \ref{sec:results} describes the general properties of the simulated PGRs and compares them with the observed PRGs. Sections \ref{sec:discussion} and \ref{sec: conclusion} contain a discussion and summary and conclusions of the main results. 

\section{Numerical methods}\label{sec: Methods}
Modern cosmological simulations have become a powerful tool for statistically studying the impact of environment and interactions on the formation and evolution of galaxies from a statistical point of view. For this reason, we used the IllustrisTNG simulation\footnote{\url{https://www.tng-project.org/}}  \citep{2018_Marinacci,2018_Naiman,2018_Nelso,2018_Pillepich,2018_Springel}, which reproduces key observational results such as the red sequence and blue cloud of SDSS galaxies \citep{Nelson_2018a}, the size evolution of galaxies \citep{Genel_2017}, the structure and kinematics of gas and stellar disks \citep{Pillepich_2019}, and the morphologies of galaxies in comparison with Pan-STARRS observations \citep{Rodriguez_Gomez_2019}, among others. 
IllustrisTNG is a magnetohydrodynamic cosmological simulation that is publicly available in all its versions, TNG50, TNG100, and TNG300, where the number in the label refers to the size of the simulation box in comoving Mpc. The simulations were calculated with the moving mesh code \texttt{AREPO} \citep{Arepo_2010} and the initial conditions were created using the \texttt{N-GENIC} code \citep{N-genic_2015} at redshift 127 using the Zeldovich approximation. The cosmological parameters used were those reported by \cite{Planck_2016}, namely $\Omega_{\rm m} = \Omega_{\rm dm} + \Omega_{\rm b} = 0.3089$, $\Omega_{\rm b}=0.0486$, and $\Omega_\Lambda = 0.6911$ (matter density, baryonic density, and cosmological constant{, respectively}), the Hubble constant used was $H_0 = 100 h$ km s$^{-1}$ with $h=0.6774$, normalisation $\sigma_8 = 0.8159$ and spectral index $n_{\rm s} = 0.9667$.

For this work, we used the TNG50-1 version \citep{Pillepich_2019, Nelson_2019} {which}, {with a computational domain of} 51.7 comoving Mpc, has a better spatial resolution than the larger versions, reaching 6.5 pc in the smallest gas cell of the simulation (at redshift $1$) and masses of baryonic particles around 0.08$\times10^{6}$ M$_\odot$ \citep{Pillepich_2019}. This allows us to better resolve the morphology of galaxies. Stars are formed stochastically in cells where the gas number density exceeds the threshold $n_{\rm thr}=0.13$ cm$^{-3}$ \citep[see][for details]{Vogelsberger_2013}. {IllustrisTNG is the continuation of the original simulation Illustris {\citep{2014_Genel,2014_Vogelsberger}} with the {goal} of reproducing with more precision properties of observed galaxies {\citep{2019a_Nelson}}. The IllustrisTNG model for galaxy formation includes star formation, stellar evolution, primordial and metal-line cooling, stellar feedback, formation, growth, and feedback of supermassive black holes \citep{2017_Weinberger, 2018a_Pillepich}. The star formation model was based on \citet{2003_Springel}, where the stars are formed stochastically after the density overpasses a density threshold. Unresolved SNe feedback was included for a star-forming gas using a two-phase, effective equation-of-state model. This galaxy formation model was calibrated on observational data to reproduce the statistical properties of galaxies. A {more detailed} description can be found in \citet{2018a_Pillepich}. }

The galaxy sample was taken from the subhalos of the snapshot corresponding to redshift zero. Subhalos were defined as a group of gravitationally bound particles that belong to a given halo and were identified using the \texttt{SUBFIND} algorithm \citep{Springel_2001}. The halos were identified using the Friends-of-Friends (FoF) algorithm introduced by \cite{Davis_1985} using a link length of $b=0.2$ times the mean particle separation. The algorithm was applied only to dark matter particles, and the other particle types were assigned to the same halo as the nearest dark matter particle. {The TNG50 {simulation} is a combination of large volumes and high resolution analogues to modern ``zoom'' simulations of individual galaxies. This simulation {generates} $\sim 20,000$ resolved galaxies with M$_\star \gtrsim 10^7$ M$_\odot$ \citep{Nelson_2019}}. 
{To ensure sufficient resolution for reliable morphological and kinematic analysis, we selected only subhalos containing at least $10^4$ stellar particles \citep{Penoyre_2017}, corresponding to a minimum stellar mass of $\sim 10^9$ M$_\odot$ (assuming a typical stellar particle mass of $10^5$ M$_\odot$). This criterion yields a sample of 4349 galaxies above the mass threshold.}

\begin{figure}
    \centering
    \includegraphics[width=\columnwidth]{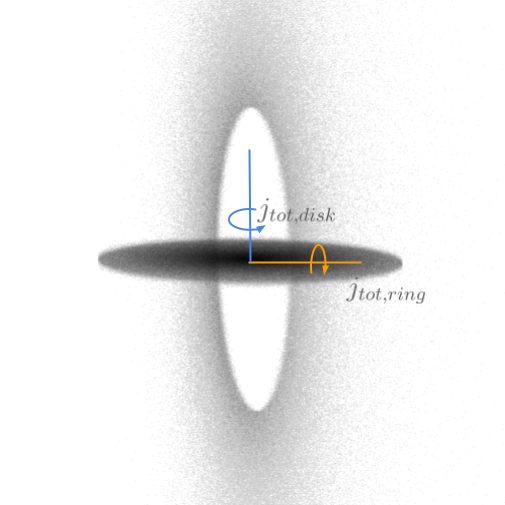}
    \caption{Representation of the vectors of the specific total angular momentum of the disk (blue) and the ring (orange) of an idealised model of a PRG.}
    \label{fig:Modelo}
\end{figure}

To identify PRGs, we use the specific angular momentum of the stars. In an ideal PRG, the total angular momentum of the host galaxy's disk and that of the ring should be perpendicular to each other (see Fig. \ref{fig:Modelo} for a schematic representation).

To define the host galaxy's reference frame, we considered the subhalo centre as the particle with the deepest gravitational potential. Additionally, particle velocities were measured in the reference frame of the centre of mass of the stellar particles within twice the subhalo stellar half-mass radius ($r_{\rm hm}$). This approach prevents the inclusion of stellar particles from satellites or unbound high-speed particles.

Once the reference frame was established, we constructed the specific angular momentum ($j$) profile of the subhalos using the particles contained within concentric spherical shells. The total specific angular momentum ($j_{\text{\rm tot}}$) of each shell was computed as:
\begin{equation}
    \textbf{j}_{\text{\rm tot}} = \dfrac{\sum m_i\textbf{r}_i \times \textbf{v}_i}{\sum m_i},
    \label{j_tot}
\end{equation}  
where $m_i$, $\textbf{r}_i$, and $\textbf{v}_i$ represent the mass, position, and velocity of each particle $i$ in the given shell. We took as a reference the $j_{\text{\rm tot}}$ of the stars within $0.5 r_{\rm hm}$ and compared its direction with the $j_{\text{\rm tot}}$ calculated in each shell. The same analysis was performed for the gaseous component, using the stellar $j_{\text{\rm tot}}$ as a reference.

Following \cite{Reshetnikov_2015}, we considered a potential PRG if the inclination angle was between 30 and 150 degrees. If the minimum inclination in the profile of the galaxy differed by more than 30 degrees and less than 150 degrees, the galaxy was considered a potential PRG as was found by \cite{Reshetnikov_2015} in their sample of observed PRG galaxies.
\begin{figure*}
    \centering
    \includegraphics[width=\textwidth]{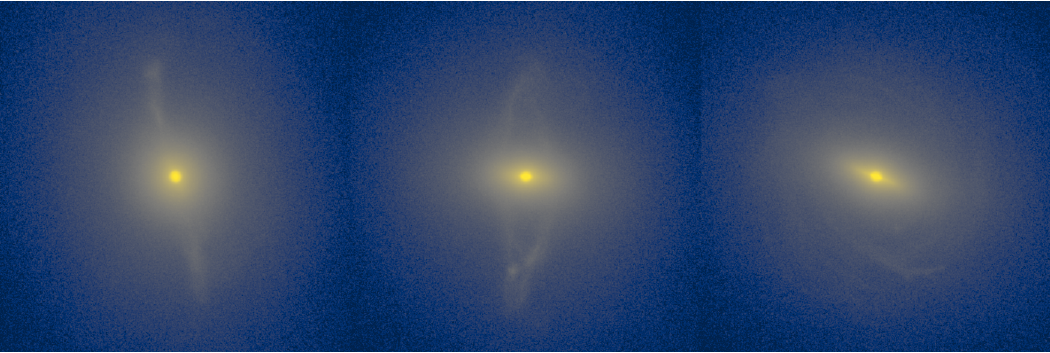} \\
    \includegraphics[trim={0.3cm 0.3cm 1.55cm 11.88cm},clip,width=\textwidth]{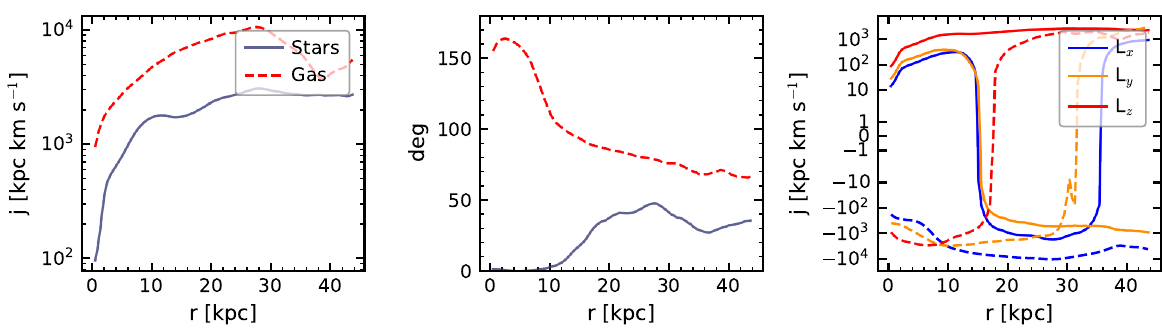}
    \caption{Stellar and gaseous structure and kinematics of a representative PRG candidate (subhalo 167392). {{\it Top row}}: surface mass density of the stellar component in the $x-y$ (left), $x-z$ (middle), and $y-z$ {planes} (right) for the subhalo 167392. {The $z$ axis was defined as the direction of the total angular momentum of the stellar particles within $0.1 r_{200}$ in the subhalo.} {{\it Bottom row}:} radial profiles of the specific angular momentum (left), the relative angle of the specific angular momentum with the $z$ axis (middle), and the components of the specific angular momentum (right). In the three panels, both the angular momentum corresponding to the stellar ({\it solid lines}) and gaseous ({\it dashed}) galaxy components are shown.}
    \label{fig:167392}
\end{figure*}

\begin{figure*}
    \centering
    \includegraphics[trim={0.6cm 5.95cm 0.2cm 6.13cm}, clip,width=\textwidth]{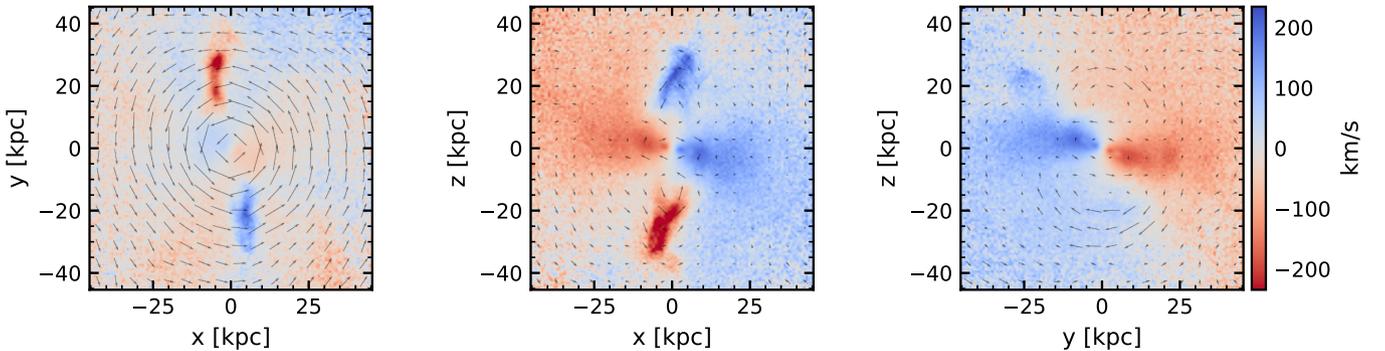}
    \caption{ Mass weighted velocities of the stellar component in the line of sight along the main axes of the subhalo 167392. The rotating ring is clearly visible as polar structures in the left and middle panels, while the rotating host disk appears as an equatorial structure in the middle and right panels.}
    \label{fig:167392_los}
\end{figure*}

Figure \ref{fig:167392} shows, as an example, the subhalo 167392 at {redshift zero}. The galaxy was reoriented by aligning $j_\text{\rm tot}$ of the stellar particles within $0.1 r_{200}$\footnote{Radius at which the mean density is 200 times the critical density of the universe} with the $z$ axis. The lower panels show the resulting radial profiles of the specific angular momentum, {its direction}, and the {magnitude of its} components. In this example, it can be seen that the ring is roughly perpendicular to the pseudo-disk. However, the maximum inclination shown in the plot is approximately 50 degrees. This difference is due to the extended bulge and stellar halos around the host galaxy. This condition complicates the selection of PRGs.
{Therefore, a} visual inspection was {also} performed using projections, profiles and line-of-sight velocity maps (see an example in Figure \ref{fig:167392_los}). 

The above described process yielded 44 PRG candidates in the entire computational box of the TNG50 simulation at redshift zero. Projections of the stellar and gaseous components of these PRGs candidates are shown in figures \ref{fig:collage_1} and \ref{fig:collage_2}. In the stellar projection, an extended spherical component is observed in all galaxies of the sample. This component makes it difficult to distinguish the ring or any other morphology. We note that the gas better traces the morphology of structures (figures \ref{fig:collage_1} and \ref{fig:collage_2}) and, interestingly, many of these galaxies exhibit a central gas deficiency, a characteristic commonly associated with late-type galaxies. 

\begin{figure*}
    \centering
    \includegraphics[trim={0.3cm 19.6cm 0.3cm 0.3cm},clip,width=0.99\textwidth]{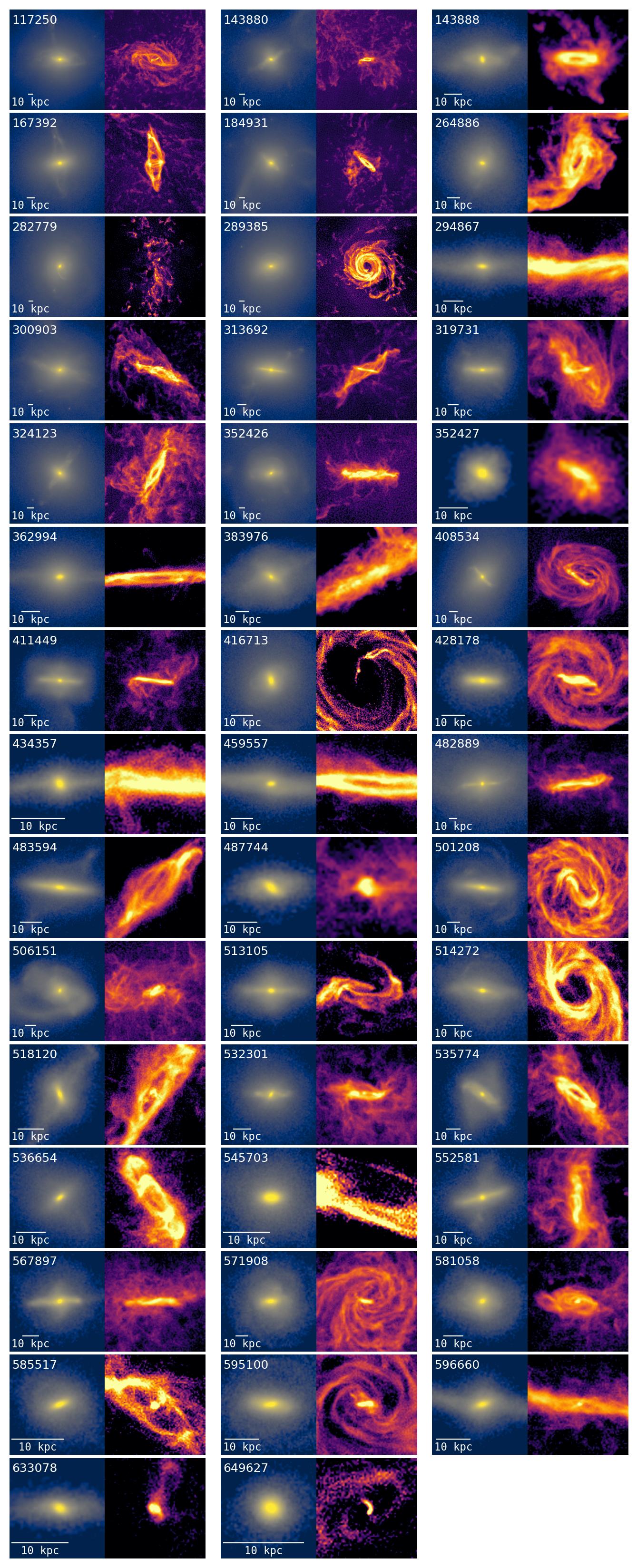}
    \caption{Surface mass density of the stellar ({\it left}) and gaseous ({\it right}) components of the first sample of galaxies listed in the table \ref{tab: kappa}}
    \label{fig:collage_1}
\end{figure*}
\begin{figure*}
    \centering
    \includegraphics[trim={0.3cm 0.3cm 0.3cm 22.4cm},clip, width=0.99\textwidth]{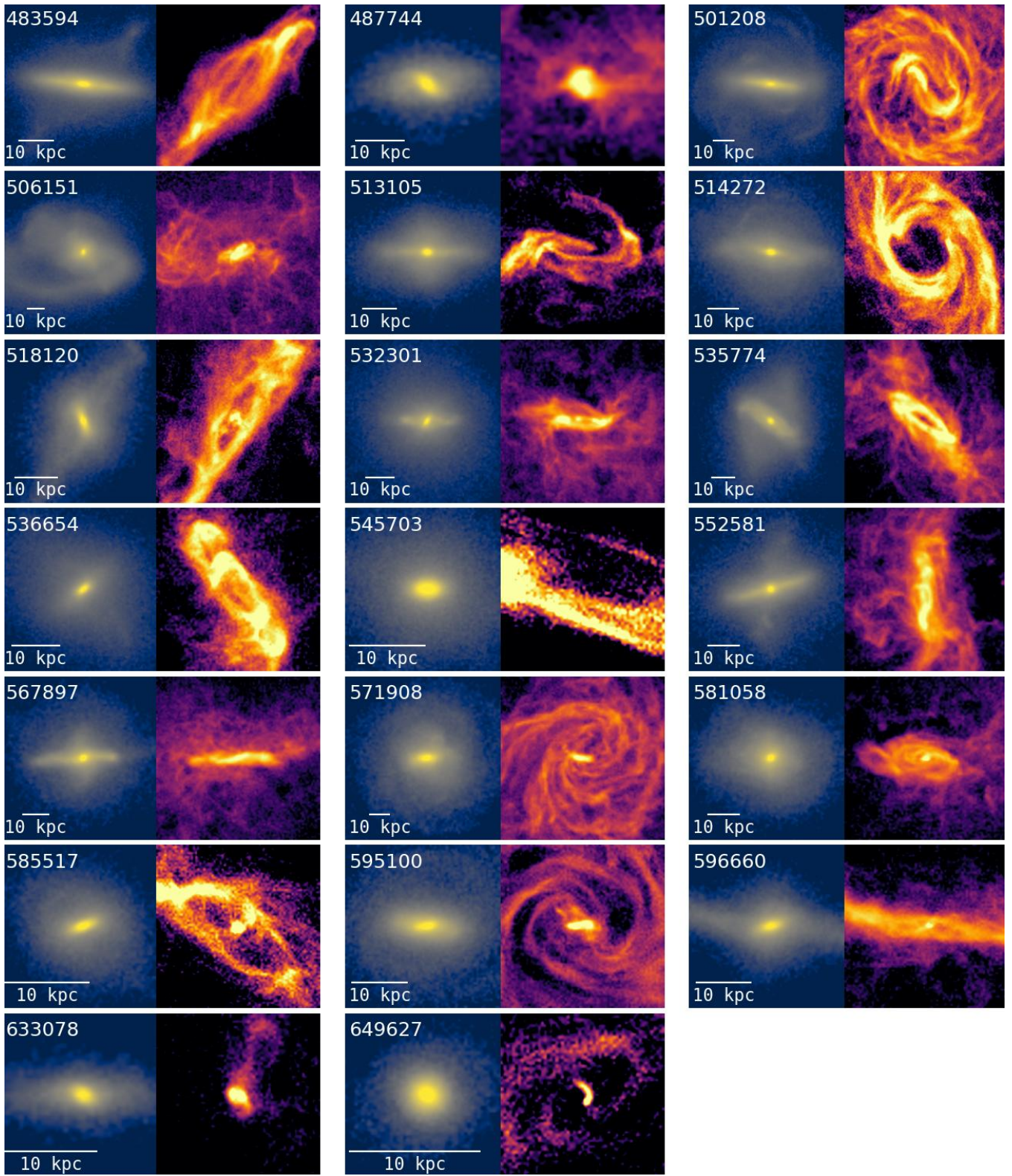}
    \caption{Continuation of Fig. \ref{fig:collage_1}}
    \label{fig:collage_2}
\end{figure*}

\subsection{Decomposition}\label{Decomposition}

To analyse the properties of our sample, it was necessary to identify the different dynamical components in galaxies. Visual inspection of the angular momentum profiles was not sufficient to identify the ring due to the extended stellar spherical component present in almost all subhalos, as shown in figures \ref{fig:collage_1} and \ref{fig:collage_2}. {Therefore, to decompose the galaxies in the sample, we used the following procedure.}
First, we examine the $\kappa$ {parameter} introduced by \cite{Sales_2012},
\begin{equation}\label{eq: kappa}
    \kappa=\dfrac{K_{\rm rot}}{K}, \hspace{1cm} K_{\rm rot}= \sum_i \dfrac{1}{2} m_i \left( \dfrac{j_{z,i}}{R_i} \right)^2,
\end{equation}
where $K_{\rm rot}$ is the total kinetic energy associated with rotation around the $z$ axis, calculated with the stellar mass ($m_i$), the specific angular momentum along the $z$ axis ($j_{z,i}$), and the cylindrical radius ($R_i$) of each particle{; and} $K$ is the total kinetic energy of the galaxy. This parameter is used as an indicator of galaxy's shape, indicating whether it is more spherical or flat. For {disk-dominated} galaxies, $\kappa {\sim 1}$, while for bulge-dominated or elliptical galaxies ${\kappa \sim} 1/3$. The values for our sample are around $1/3$, except for four of them that have values larger than 0.5 (see {Table} \ref{tab: kappa}). This parameter classifies most of the galaxies in our sample as spherical or elliptical galaxies, due to the extended spherical component, as shown in figures \ref{fig:collage_1} and \ref{fig:collage_2}. 

\begin{table}
\caption{(1) ID of the subhalos in TNG50. (2) $\kappa$ parameter described by eq. \ref{eq: kappa}. (3) Mass fraction of particles with $\epsilon>0.7$. (4) Mass fraction of particles with $\epsilon < -0.7$  }
\label{tab: kappa}
\begin{tabular}{cccc}
\multicolumn{1}{l}{Subhalo ID$^{(1)}$} & $\kappa^{(2)}$ & $\epsilon_{07}^{(3)}$ & $\epsilon_{-07}^{(4)}$ \\ \hline
117250                         & 0.298    & 0.199         & 0.036          \\
143880                         & 0.334    & 0.133         & 0.057          \\
143888                         & 0.418    & 0.256         & 0.064          \\
167392                         & 0.404    & 0.074         & 0.059          \\
184931                         & 0.300    & 0.074         & 0.061          \\
264886                         & 0.289    & 0.067         & 0.055          \\
282779                         & 0.290    & 0.080         & 0.046          \\
289385                         & 0.330    & 0.162         & 0.039          \\
294867                         & 0.339    & 0.226         & 0.073          \\
300903                         & 0.382    & 0.193         & 0.041          \\
313692                         & 0.388    & 0.335         & 0.033          \\
319731                         & 0.411    & 0.462         & 0.016          \\
324123                         & 0.338    & 0.117         & 0.044          \\
352426                         & 0.338    & 0.169         & 0.030          \\
352427                         & 0.361    & 0.163         & 0.032          \\
362994                         & 0.350    & 0.129         & 0.062          \\
383976                         & 0.395    & 0.200         & 0.026          \\
408534                         & 0.323    & 0.080         & 0.039          \\
411449                         & 0.330    & 0.285         & 0.030          \\
416713                         & 0.361    & 0.051         & 0.050          \\
428178                         & 0.562    & 0.549         & 0.010          \\
434357                         & 0.451    & 0.206         & 0.067          \\
459557                         & 0.261    & 0.229         & 0.139          \\
482889                         & 0.379    & 0.165         & 0.056          \\
483594                         & 0.217    & 0.389         & 0.023          \\
487744                         & 0.301    & 0.268         & 0.035          \\
501208                         & 0.606    & 0.509         & 0.014          \\
506151                         & 0.410    & 0.173         & 0.022          \\
513105                         & 0.363    & 0.222         & 0.088          \\
514272                         & 0.289    & 0.133         & 0.023          \\
518120                         & 0.317    & 0.051         & 0.065          \\
532301                         & 0.335    & 0.192         & 0.032          \\
535774                         & 0.457    & 0.072         & 0.085          \\
536654                         & 0.377    & 0.113         & 0.033          \\
545703                         & 0.302    & 0.092         & 0.053          \\
552581                         & 0.373    & 0.344         & 0.039          \\
567897                         & 0.445    & 0.348         & 0.025          \\
571908                         & 0.543    & 0.305         & 0.016          \\
581058                         & 0.401    & 0.163         & 0.048          \\
585517                         & 0.316    & 0.206         & 0.024          \\
595100                         & 0.637    & 0.436         & 0.008          \\
596660                         & 0.309    & 0.364         & 0.030          \\
633078                         & 0.255    & 0.245         & 0.093          \\
649627                         & 0.360    & 0.157         & 0.052          \\ \hline
\end{tabular}
\end{table}

\begin{figure*}
    \centering
    \includegraphics[trim={0.41cm 0.44cm 0.37cm 0.35cm},clip,width=\textwidth]{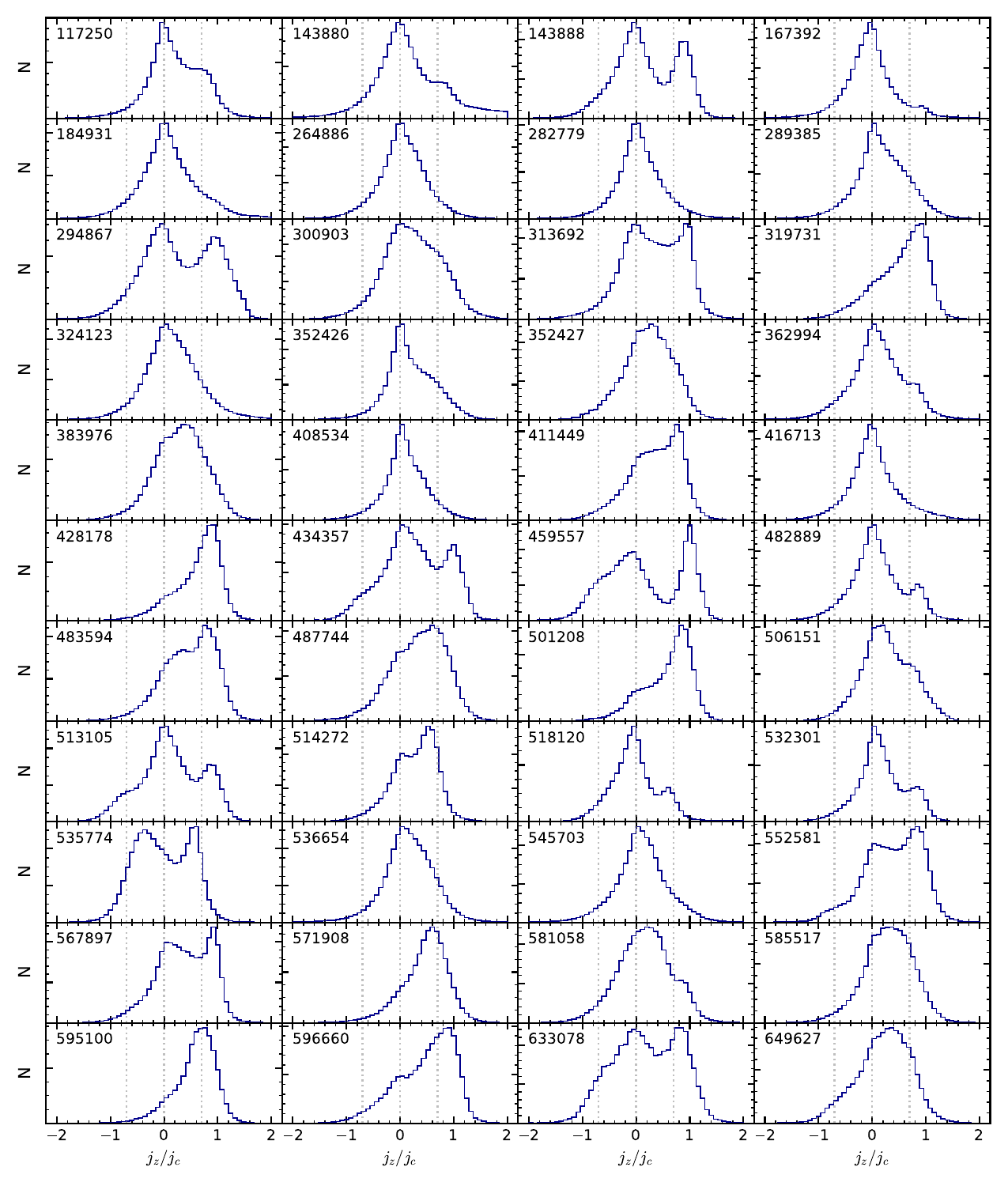}
    \caption{Circularities distribution of the complete first sample {from} table \ref{tab: kappa}{. For each panel, the subhalo ID is indicated as the number in the upper left corner.} Dotted vertical lines correspond to values of -0.7, 0, and 0.7 respectively.}
    \label{fig:circularity}
\end{figure*}

Circularity ($\epsilon$) is a widely used parameter in the literature on dynamical decomposition, introduced by \citet{Abadi_2003}. In this work, we used the version of \citet{Scannapieco_2009}\footnote{ Both definitions yield very similar results \citep{Marinacci_2013}.} which is defined as
\begin{equation}
    \epsilon = \dfrac{\textbf{j} \cdot \hat{\textbf{z}} }{j_{\rm c}(r)},
\end{equation}
{where}
\begin{equation}
    j_{\rm c}(r) = rv_{\rm c}(r) = r \sqrt{\dfrac{GM(r)}{r}}
\end{equation}
{is the angular momentum associated with a circular orbit.}
Values of $|\epsilon|$ greater than 0.7 correspond to the particles in a disk, with negative $\epsilon$ values corresponding to a counter-rotating disk. Particles {with $|\epsilon| < 0.7$ belong} to spherical components. Figure \ref{fig:circularity} shows the {distribution of particle} circularities in the subhalos of our sample. These distributions show that almost all galaxies are dominated by a spherical component except for {the four} galaxies with $\kappa \sim 0.6$ (IDs: 428178, 501208, 571908, and 595100) that exhibit right-skewed peaks. The remaining subhalos show prominent peaks around $\epsilon \sim 0$, indicating the dominance of the spherical components in the systems.

\begin{figure}
    \centering
    \includegraphics[trim={0.1cm 0.05cm 1.11cm 1.05cm},clip, width=\columnwidth]{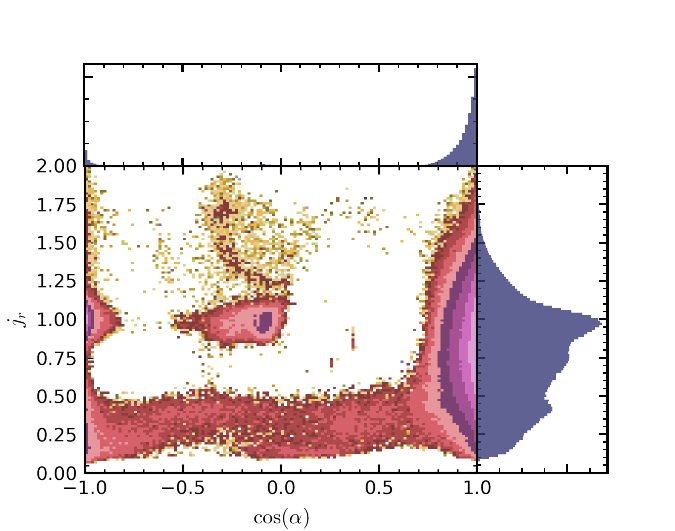}
    \caption{$\cos{\alpha}-j_{\rm r}$ {diagram} of the subhalo 167392. The colours indicate the density of particles. The upper and right histograms are the projections of the diagram along the $\cos{\alpha}$ and $j_{\rm r}$ axes.}
    \label{fig:jr-alpha}
\end{figure}

The values obtained for $\kappa$ and $\epsilon$ are consistent at different levels, but so far none of these provide much information about the rings. {In order} to separate the rings from the host galaxy, a variation of the method proposed by \citet{jagvaral_probabilistic_2021} was used. This method consists in creating a $\cos(\alpha)-j_{\rm r}$ diagram, where $\alpha$ is the angle subtended between the particle angular momentum vector and the total angular momentum of the stellar particles within $1 r_{\rm hm}${, and} 
$j_{\rm r}$ is the ratio of the specific angular momentum of the particles ($j_i$) to its expected specific angular momentum for a circular orbit ($j_{\rm c}$) at the same galactocentric distance, thus describing how circular the orbit of the particle is.
{Particles that fall in one of three categories are of particular interest:} {\it i)} particles with values of $\cos{(\alpha)} \sim 1${, which} are commonly associated with a disk; {\it ii)} particles with values of $\cos{(\alpha)} \sim -1${, which} indicate a counter-rotating disk; and {\it iii)} particles with $\cos{(\alpha)} \sim 0${, which} indicate structures orbiting perpendicular {to the plane of the galaxy}, suggesting the presence of a polar ring. Figure \ref{fig:jr-alpha} shows the $\cos(\alpha)-j_{\rm r}$ {diagram} for the subhalo 167392. Different groups are visible in reddish colours: a disk on the right ($\cos{(\alpha)} \sim 1$ {and $j_{\rm r} \sim 0.75$}), a ring at the centre ($\cos{(\alpha)} \sim 0${, $j_{\rm r} \sim 1$}), and a counter-rotating component to the left ($\cos{(\alpha)} \sim -1${, $j_{\rm r} \sim 1$}).

\begin{figure}
    \centering
    \includegraphics[trim={0.2cm 0.4cm 0.3cm 0.7cm},clip,width=\columnwidth]{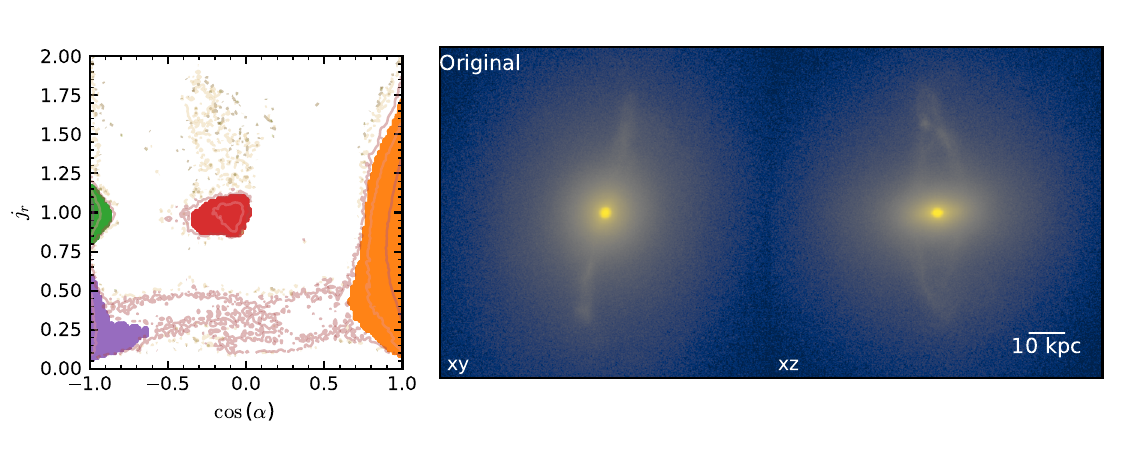}
    \includegraphics[trim={0.45cm 0.3cm 0.45cm 0.3cm},clip,width=\columnwidth]{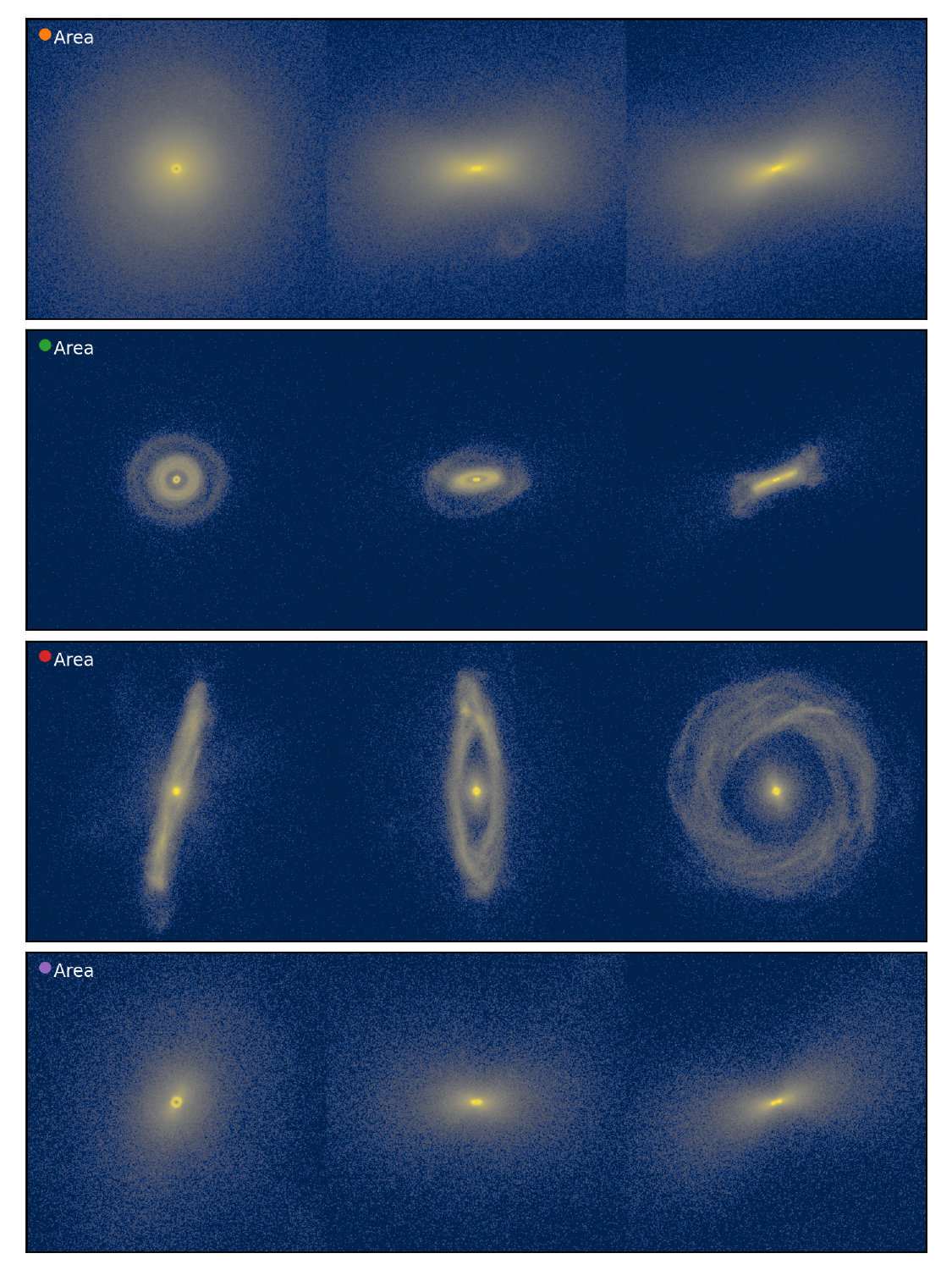}
    \caption{Decomposition of the subhalo 167392. The upper left panel shows the {$\cos{\alpha}-j_{\rm r}$ diagram, with the coloured areas indicating the individual components of the subhalo}. The upper right panel shows the original subhalo before the decomposition. The remaining panels display the projections of the stellar particles that fall within the shaded regions of the $\cos{\alpha}-j_{\rm r}$ diagram. Each component is marked by a coloured dot in the upper left corner of its respective panel: orange for the host galaxy disk, green and purple for the counter-rotating disk component, and red for the polar ring.}
    \label{fig:decomposition}
\end{figure}

To separate the particles of each component {in the} $\cos{\alpha}-j_{\rm r}$ {diagram}, we used {the {\tt CUPID}\footnote{\url{http://starlink.eao.hawaii.edu/starlink/CUPID}}} clump finding algorithm \citep{Cupid_Berry}. First, the background was removed from the $\cos{\alpha}-j_{\rm r}$ diagram using a method similar to that used for clumpiness \citep{2003_Conselice}.  
It was removed by smoothing the original image with a median filter, where the filter size was set to half the square root of the number of bins and then subtracted from the original image.
Finally, the clump finding algorithm was applied to the background-subtracted image. The method was initially applied to the stellar particles as these provide a better sample of the dynamic behaviour exhibited by the different galactic components. It was then repeated for the gaseous component, with the orientation of the stellar disk serving as the reference {direction}. Figure \ref{fig:decomposition} shows the different components obtained by this method for the stellar component. The orange area in the $\cos{\alpha}-j_{\rm r}$ diagram corresponds to the main disk of the host galaxy. The components depicted by the green and purple colours correspond to a counter-rotating disk, the first with very circular orbits and the second with more dispersion. While these three regions are coplanar, a fourth structure is visible near the centre of the $\cos{\alpha}-j_{\rm r}$ diagram. {This area, marked in red colour, corresponds to a} ring {comprised of particles in} very circular orbits, in contrast to the host galaxy, which shows a larger dispersion similar to ETGs. {The individual components are also shown in figure \ref{fig:decomposition}.}

Dynamic decomposition allows for a better understanding of the different components and their relation to counter-rotating disks and PRGs \citep{2020_Khoperskov}. This method reveals the different dynamical components and allows the identification and classification of structures in the subhalos. From the initial sample of 44 galaxies, 12 were excluded (143880, 184931, 282779, 294867, 300903, 352426, 362994, 459557, 482889, 506151, 567897, and 581058), as the decomposition did not show a clear polar structure in these subhalos. Some of these galaxies have structures resembling a merger remnant, while others, such as subhalo 362994, resemble a ring around an extended bulge, but not in a polar orientation (similar to the Sombrero galaxy). {As shown in Figures \ref{fig:collage_1} and \ref{fig:collage_2}, galaxies in this sample often exhibit extended spherical components or tidal streams, likely associated with violent formation events or past interactions. These structures add noise to the analysis, which may obscure faint polar features or even lead to misidentifications. Particles in stellar halos or interaction remnants that have a similar dynamic as rings may produce spurious measurements. Such contamination could artificially increase the number of PRG candidates or, conversely, hide genuine faint polar structures. Although our method aims to minimise this contamination and incorporates multiple filters, we acknowledge that residual non-rotational components may still bias the estimated properties of both the rings and their host galaxies.}

\begin{table*}
\centering
\caption{Total and component masses of the PRGs in TNG50. All the values were calculated within 5 $r_{\rm hm}$}
\label{Masses}
\begin{tabular}{S[table-format=6]S[table-format=2.3]S[table-format=2.3]S[table-format=3.3]S[table-format=1.3]S[table-format=2.3]S[table-format=1.3]}
\multicolumn{1}{c}{Subhalo ID} & \multicolumn{1}{c}{\begin{tabular}[c]{@{}c@{}}Stellar Mass\\{[}$10^{10}$ M$_\odot$]\end{tabular}} & \multicolumn{1}{c}{\begin{tabular}[c]{@{}c@{}}Gas Mass\\{[}$10^{10}$ M$_\odot$]\end{tabular}} & \multicolumn{1}{c}{\begin{tabular}[c]{@{}c@{}}Baryonic mass \\{[}$10^{10}$ M$_\odot$]\end{tabular}} & \multicolumn{1}{c}{\begin{tabular}[c]{@{}c@{}}Stellar ring mass\\{[}$10^{10}$ M$_\odot$]\end{tabular}} & \multicolumn{1}{c}{\begin{tabular}[c]{@{}c@{}}Stellar host mass\\{[}$10^{10}$ M$_\odot$]\end{tabular}} & \multicolumn{1}{c}{\begin{tabular}[c]{@{}c@{}}Ring to host\\~mass ratio\end{tabular}} \\ \hline
117250 & 97.835 & 19.385 & 117.220 & 3.740 & 94.216 & 0.040 \\
143888 & 9.544 & 0.711 & 10.255 & 2.681 & 7.232 & 0.371 \\
167392 & 44.133 & 5.955 & 50.088 & 1.075 & 43.127 & 0.025 \\
264886 & 7.911 & 2.093 & 10.004 & 0.806 & 7.135 & 0.113 \\
289385 & 63.499 & 5.119 & 68.617 & 1.248 & 62.291 & 0.020 \\
313692 & 43.780 & 6.769 & 50.548 & 0.545 & 43.250 & 0.013 \\
319731 & 7.167 & 2.483 & 9.650 & 0.256 & 6.924 & 0.037 \\
324123 & 29.857 & 8.790 & 38.647 & 1.667 & 28.303 & 0.059 \\
352427 & 0.198 & 0.140 & 0.338 & 0.027 & 0.171 & 0.160 \\
383976 & 6.745 & 2.856 & 9.601 & 0.449 & 6.327 & 0.071 \\
408534 & 17.409 & 1.993 & 19.402 & 0.912 & 16.535 & 0.055 \\
411449 & 15.673 & 0.912 & 16.586 & 0.918 & 14.801 & 0.062 \\
416713 & 11.149 & 0.218 & 11.367 & 0.550 & 10.833 & 0.051 \\
428178 & 1.867 & 0.717 & 2.584 & 0.081 & 1.795 & 0.045 \\
434357 & 0.832 & 0.254 & 1.086 & 0.254 & 0.683 & 0.371 \\
483594 & 8.606 & 0.703 & 9.309 & 0.540 & 8.178 & 0.066 \\
487744 & 0.388 & 0.175 & 0.563 & 0.060 & 0.331 & 0.182 \\
501208 & 8.355 & 2.378 & 10.733 & 0.331 & 8.056 & 0.041 \\
513105 & 4.335 & 0.219 & 4.553 & 0.538 & 3.801 & 0.141 \\
514272 & 3.503 & 0.898 & 4.401 & 0.123 & 3.388 & 0.036 \\
518120 & 1.795 & 0.841 & 2.636 & 0.233 & 1.637 & 0.142 \\
532301 & 4.750 & 1.852 & 6.601 & 0.289 & 4.551 & 0.063 \\
535774 & 4.921 & 2.710 & 7.631 & 0.365 & 4.586 & 0.080 \\
536654 & 3.172 & 0.488 & 3.661 & 0.211 & 3.003 & 0.070 \\
545703 & 2.705 & 0.004 & 2.710 & 0.191 & 2.546 & 0.075 \\
552581 & 4.384 & 1.299 & 5.684 & 0.124 & 4.262 & 0.029 \\
571908 & 2.381 & 2.031 & 4.412 & 0.358 & 2.076 & 0.172 \\
585517 & 1.238 & 0.161 & 1.399 & 0.136 & 1.119 & 0.122 \\
595100 & 1.969 & 0.420 & 2.389 & 0.145 & 1.880 & 0.077 \\
596660 & 1.453 & 0.297 & 1.750 & 0.205 & 1.321 & 0.155 \\
633078 & 0.270 & 0.060 & 0.329 & 0.097 & 0.190 & 0.510 \\
649627 & 0.757 & 0.041 & 0.798 & 0.061 & 0.713 & 0.086 \\ \hline
\end{tabular}
\end{table*}

\section{Results}\label{sec:results}

The final sample of PRG in the TNG50 simulation is conformed by 32 subhalos. The decomposition allows for the selection of the particles that belong to individual structures in the subhalos. This decomposition method was also applied to the gaseous component of the galaxy. Due to the lack of gas in the inner part of many subhalos in the sample, we take as reference the total angular momentum of the stellar particles inside $1 r_{\rm hm}$. The PRG sample in TNG50 is listed in table \ref{Masses} with the masses of the stellar and gaseous components in the whole galaxy within a radius of $5 r_{\rm hm}$.

Individual and statistical analyses have been performed on PRGs since they were first observed \citep{1991_Whitmore,2002_Iodice,Reshetnikov_2015}. Taking advantage of the number of galaxies from cosmological simulations, it is possible to compare the similarity of our sample against the observed properties. In this work, we compared our sample of simulated PRG with the observational PRG sample of \citet{Reshetnikov_2015}, composed of 46 objects from the SDSS-based Polar Ring Catalogue (SPRC) \citep{2011_Moiseev} and 4 from the catalogue presented in \citet{1990_Whitmore}.

\subsection{Fraction of galaxies with polar ring}

PRGs are uncommon objects. \citet{1990_Whitmore} estimated that 5\% of near{by} lenticular galaxies have or had some polar structure. \citet{2011_Reshetnikov} estimated that $\sim$0.4\% of nearby galaxies, with absolute magnitudes in the range of -17 to -22 mag in the B band, have polar-ring features. Also, \citet{2022_Smirnov} found $\sim 0.01\%$ of PRG in a sample of galaxies in the $r$ band with $M_r =$ -17 to -22 mag, concluding that this fraction increases with redshift up to $ \sim 1$. Recently, \citet{2024_Mosenkov} reported that the fraction of PRG in the $r$-band is $\sim 1.1\%$ or $\sim  3\%$ when projection effects are taken into account.

To compare our sample with observations, we used the synthetic multi-wavelength photometry catalogue presented by \citet{2022_Trcka}. This catalogue was produced using data from the TNG50 simulation and the radiative transfer code SKIRT \citep{2020_SKIRT}. It contains subhalos magnitudes for 14 broadband filters, ranging from UV to submillimeter wavelengths. For a direct comparison, we used the Johnson\_B and SDSS\_r filters. The fraction of galaxies with a polar ring in the TNG50 was calculated using the mean magnitude of the subhalos in the three orientations within $5 r_{\rm hm}$ to account for any orientation-related issues. For the range of $M_{\rm B}$ = -17 to -22 mag, the fraction is $1.01 \pm 0.20\%$ and for the range of M$_r$ = -17 to -22, it is 0.53$\pm$0.11\% (bootstrap estimated errors). The results obtained in TNG50 are approximately one order of magnitude larger than those reported by \citet{2011_Reshetnikov} and \citet{2022_Smirnov}, but smaller than the result obtained by \citet{2024_Mosenkov}. However, the fraction of PRGs in TNG50 filtered by stellar mass within $5 r_{\rm hm}$ between $10^{9}-10^{12}$ is 1\%.  

\begin{figure}
    \centering
    \includegraphics[width=\columnwidth]{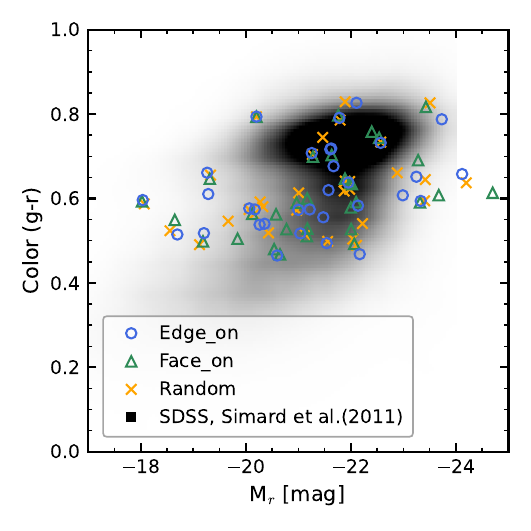}
    \caption{Colour-magnitude diagram {for galaxies on this catalogue}. The greyscale background represents the point density of galaxies taken from the Legacy area of the Sloan Digital Sky Survey Data Release Seven taken from the catalogue of \citet[][with darker areas indicating higher concentrations]{2011_Simard}. Markers are the values of the TNG50 PRGs. Magnitudes {are} taken from the catalogue of \citet{2022_Trcka} in three orientations inside $5 r_{\rm hm}$: edge-on (blue circles), face-on (green triangles), and random orientation (yellow crosses). }
    \label{fig:Color_mag}  
\end{figure}

Using the synthetic catalogue from \citet{2022_Trcka}, we plotted our PRGs in the colour-magnitude diagram (see Fig. \ref{fig:Color_mag}). The three orientations reported by \citet{2022_Trcka} {are shown} for comparison (face-on, edge-on, and randomly oriented).\footnote{Note, however, that aligning a PRG solely based on its angular momentum is imprecise due to contributions from both the host galaxy and the polar ring.} We compared the position of our PRG sample with galaxies from the Legacy area of the Sloan Digital Sky Survey Data Release Seven reported by \citet{2011_Simard}. As expected, while some PRGs lie in the red sequence, most of them are scattered in the Green Valley and the upper region of the blue cloud. 

\subsection{Bulge-to-total mass ratio in host galaxies}

The bulge-to-total mass ratio (B/T) is a widely used parameter for classifying galaxy morphology. Galaxies having $\text{B/T} \geq 0.5$ are classified as early type and late type otherwise. To estimate the bulge mass, we used a criterion similar to that described in the catalogue by \citet{2015_Genel}, where the circularity parameter described in \S \ref{Decomposition} is used. For this analysis, the parameter $\epsilon$ was calculated excluding the particles associated with the ring to minimise their influence on the results. {We estimated} the bulge {mass} using particles with $\epsilon<0$, multiplied by 2, assuming that the bulge symmetry is around 0 in the circularity plot. This helps minimise the contribution of particles in the disk. 
 
\begin{figure}
    \centering
    \includegraphics[width=\columnwidth]{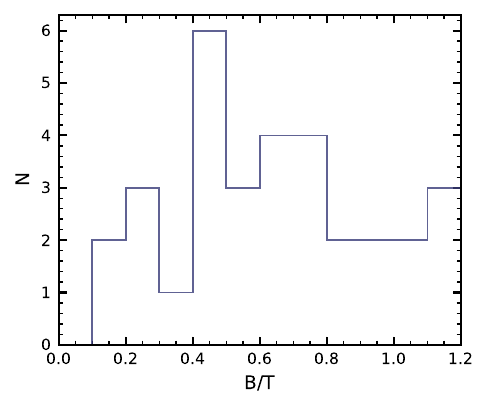}
    \caption{Bulge-to-total mass ratio (B/T) histogram of the PRG galaxies using circularities in the range $-0.7<\epsilon<0.7$ and the mass inside $5 r_{\rm hm}$.}
    \label{fig:B_T}
\end{figure}

 The results must be interpreted with caution, given the presence of counter-rotating components in our PRG sample. Although these components are not significant in most galaxies (see table \ref{tab: kappa}), they are prominent in some, such as in the 459557 subhalo. 
Our analysis shows that most host galaxies have B/T values greater than 0.5, with an average of 0.64, indicating that PRG in the TNG50 simulation are predominantly bulge-dominated systems (see Fig. \ref{fig:B_T}). Subhalos with B/T values below 0.5, such as 428178, 501208, and 595100, were also classified as disk-dominated galaxies based on the $\kappa$ parameter and the original $\epsilon$ values prior to decomposition. These results suggest a diverse range of morphologies within the PRG sample. 
 
\subsection{Ring-to-host mass ratio}

Observations have shown that the baryonic mass of the ring in PRGs is not negligible, sometimes comparable to or even exceeding the mass of the host galaxy \citep{2002_Iodice,Reshetnikov_2015}. In this study, since the subhalos have been dynamically decomposed into their individual components, estimating the masses of the ring and host galaxy was straightforward. The mass of each was calculated by summing the mass of the stellar particles assigned to either the ring or the host galaxy. Ring particles were identified as those falling within the central shaded regions in the $\cos{\alpha}-j_{\rm r}$ {diagram}, while the host galaxy consists of the remaining stellar particles. These masses were computed for all stellar particles within a radius of $5 r_{\rm hm}$. 

\begin{figure}
    \centering
    \includegraphics[width=\columnwidth]{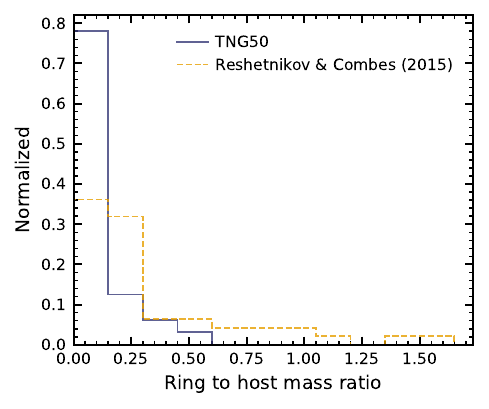}
    \caption{Ring to host mass ratio for the TNG50 PRG sample (solid line) and the observed PRG (dashed line) from \citet{Reshetnikov_2015}.}
    \label{fig:Ring to host}
\end{figure}

From the distribution of the ring-to-host mass ratios in Figure \ref{fig:Ring to host}, it is clear that the rings in our TNG50 sample are less massive than those observed in real PRGs. Most of the subhalos in the sample exhibit ring-to-host mass ratios below 0.2 (none exceed 0.6). The mean value of this distribution is $\sim 0.11$, indicating that in the TNG50 simulation, rings are typically much less dominant in terms of mass compared to the host galaxy.

\subsection{Ring inclinations}

By definition, polar rings are expected to have an inclination close to 90 degrees from the plane of the host galaxy. However, observations have shown a wide range of inclinations, with many rings nearly perpendicular to the host galaxy. Some cases have reported ring inclinations as low as $\sim 35 \text{deg}$ from the plane of the host galaxy \citep{1991_Whitmore, Reshetnikov_2015}. Simulations have also shown that {ring} inclinations can change over time \citep{2023_Smirnov}. 

\begin{figure}
    \centering  
    \includegraphics[width=\columnwidth]{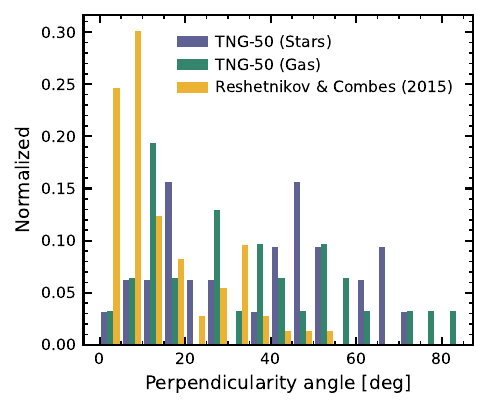}
    \includegraphics[width=\columnwidth]{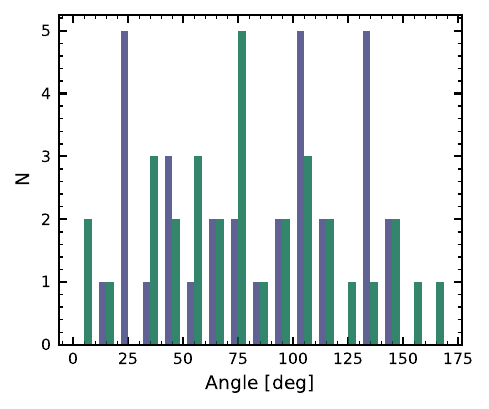}
    \caption{Angular distribution of the ring–disk orientation in the PRG sample. {{\it Top}:} Perpendicularity of the ring measured from the disk plane, shown separately for the stellar (blue) and gaseous (green) components. The yellow histogram is the perpendicularity estimated from observations of \citet{1991_Whitmore} and \citet{Reshetnikov_2015}. {{\it Bottom}:} Distribution of the estimated angles between the angular momentum vector of the stellar disk and the stellar (blue) and gaseous (green) ring.
    }
    \label{fig:perp_angles}
\end{figure}

In this work, we computed the ring inclination by comparing the angular momentum vectors of the host galaxy and the selected ring. The top panel of figure \ref{fig:perp_angles} shows the perpendicularity of stellar and gas rings, i.e., how far the components are from perpendicularity, where 0 degrees indicates an orthogonal ring and 90 degrees is a coplanar component with respect to the host galaxy. For the stellar component, there is no clear trend; the distribution shows two peaks at bins 15-20 and 45-50 degrees (blue bars), while gas rings show a trend with a peak at bin 10-15 degrees (green bars). This tendency for gas rings is more similar to the perpendicularity of observed PRGs. Angles measured from host disks show the distribution of rings in more detail (bottom panel, Fig. \ref{fig:perp_angles}). No clear pattern is observed in the angles of stellar rings, whereas gas rings show a trend centred on the 70-80 degree bin. The lower panel of the figure \ref{fig:perp_angles} also shows that two gas disks are almost antiparallel to their corresponding stellar disks, which may suggest a link between the evolution of PRGs and counter-rotating galaxies \citep{2020_Khoperskov}. Differences in the angles of the stellar and gaseous rings could be due to dynamical evolution and perturbations outside the subhalos.  

The discrepancy between the observed inclinations and our results may be due to observational biases, such as selection effects, projection angles, or resolution limits. The dynamical decomposition of the galaxies allows us to identify those with inclinations with respect to the disk. Figure \ref{fig:perp_angles} shows the distribution of the angles between the selected rings and the dominant disks. The mean inclination for the stellar rings is $\sim 82.61$ degrees, while $\sim 50\%$ of the objects have inclinations greater than 90 degrees and $\sim 50\%$ less than 90 degrees. This suggests a relatively uniform distribution of the angles and their evolution. For gaseous rings, $\sim 0.56\%$ have inclinations less than 90 degrees, with an average inclination of 80.44 degrees.  

\subsection{Ring radii}

\begin{figure}
    \centering
    \includegraphics{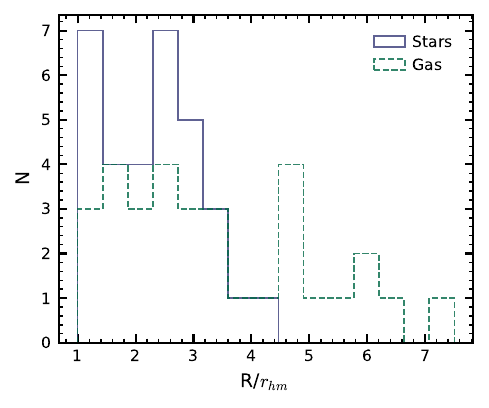}
    \caption{Distribution of the mean {ring} radius over the $r_{\rm hm}$ of the PRG sample.}
    \label{fig:mean R}
\end{figure}

The mean radius ($R$) of the rings was calculated using: 
\begin{equation}
    R = \dfrac{ \displaystyle \int_{0.5 r_{\rm hm}}^{10 r_{\rm hm}} r^2 \rho(r) dr } { \displaystyle \int_{0.5 r_{\rm hm}}^{10 r_{\rm hm}} r \rho(r) dr},
\end{equation}
as described in \citet{2023_Smirnov}, where $\rho(r)$ is the density profile of the stellar ring. The radii of the rings vary significantly depending on the size of the host galaxies, with an average radius of $\sim 10.19$ kpc in this sample. To put this in perspective, we compared the mean radius of the rings with the stellar half-mass radius of the subhalo ($r_{\rm hm}$). Most of the rings are extended, between $\sim 2 r_{\rm hm}$ and $\sim 4 r_{\rm hm}$ (see Fig. \ref{fig:mean R} and Table \ref{Results}). However, a considerable number of PRGs have ring radii smaller than $2 r_{\rm hm}$, indicating that some rings could be embedded within the central structures of their host galaxies. Additionally, it is worth noting that the rings in the TNG50 simulation appear to be thick. 

\begin{figure}
    \centering
    \includegraphics[width=0.49\columnwidth]{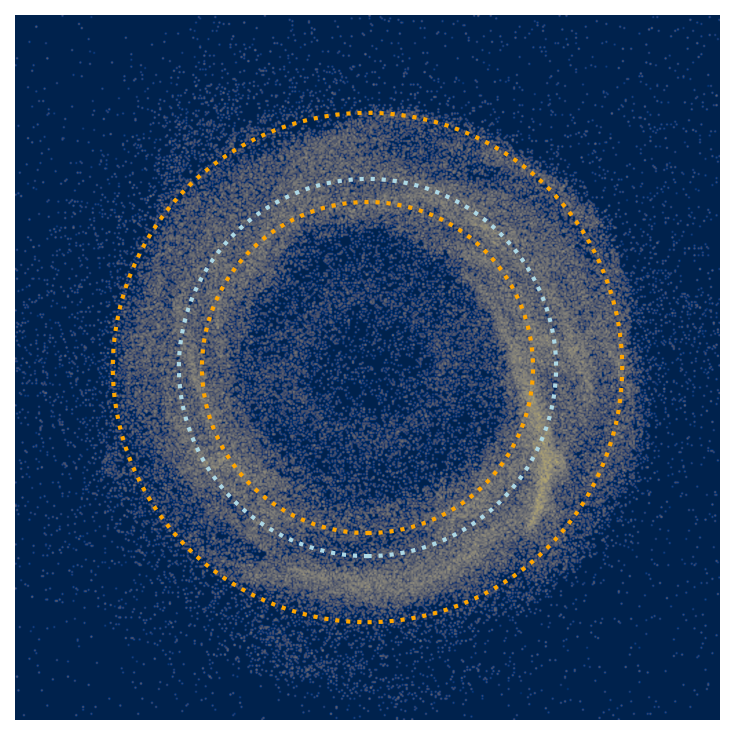}
    \includegraphics[width=0.49\columnwidth]{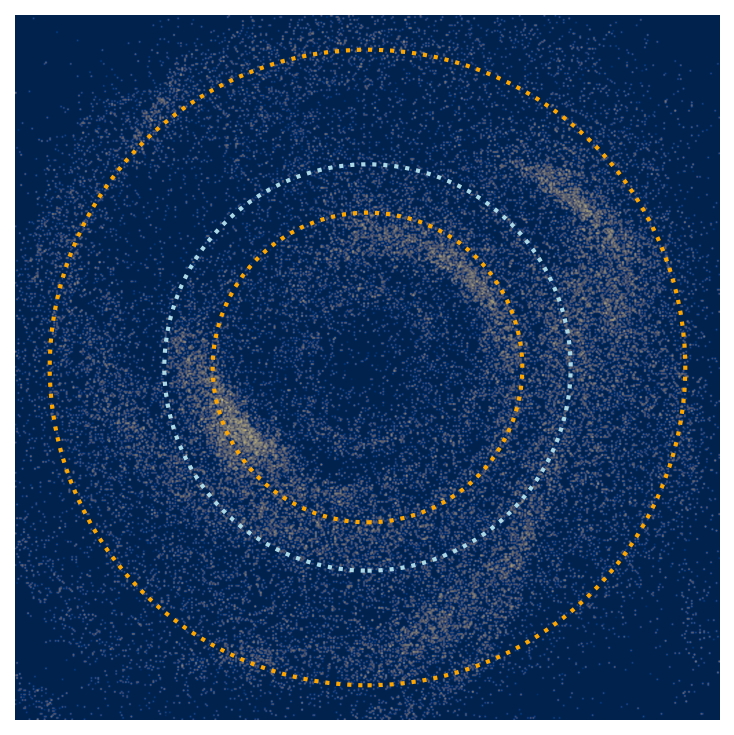}
    \includegraphics[width=0.49\columnwidth]{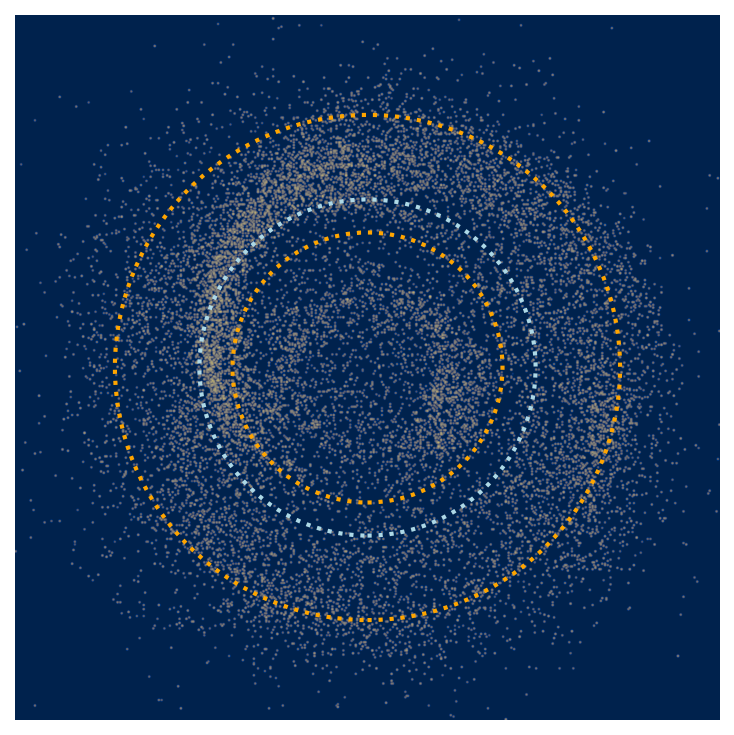}
    \includegraphics[width=0.49\columnwidth]{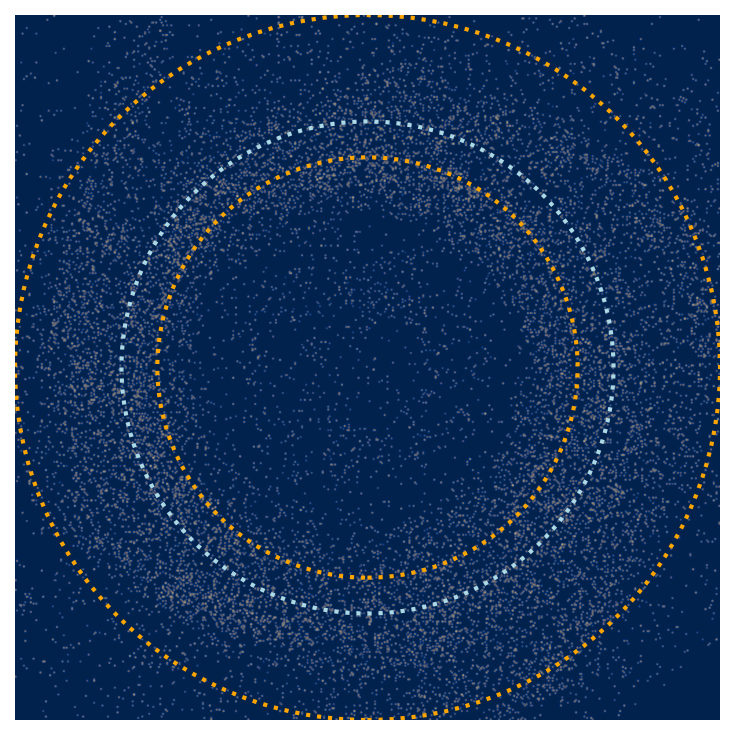}
    \caption{Example images showing the mean radius (blue dotted circle) and 20th and 80th percentiles (yellow dotted circles) of the stellar ring's surface mass density for subhalos 167392 ({\it upper left}), 501208 ({\it upper right}), 552581 ({\it lower left}), and 595100 ({\it lower right}).}
    \label{fig:percentiles}
\end{figure}

\begin{figure}
    \centering
    \includegraphics[width=0.49\columnwidth]{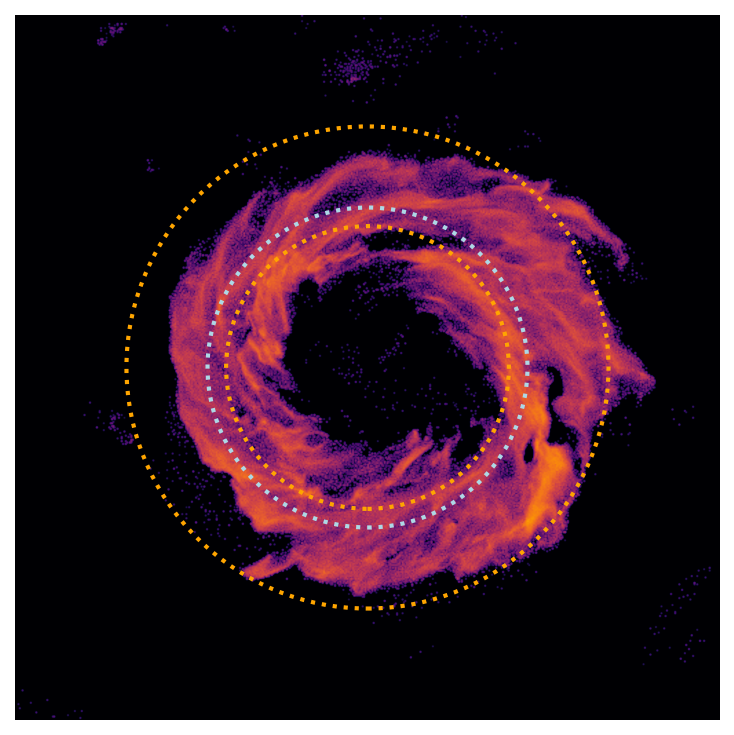}
    \includegraphics[width=0.49\columnwidth]{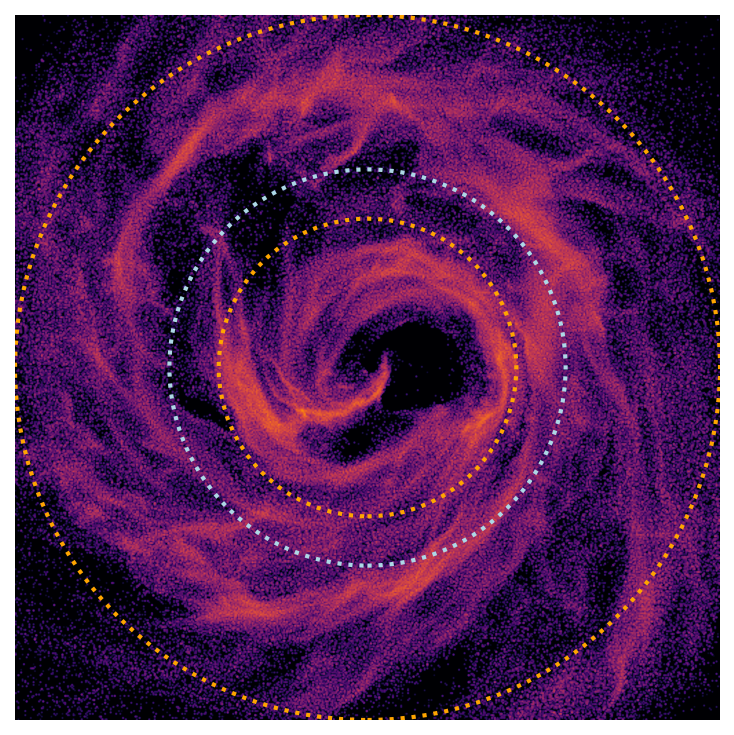}
    \includegraphics[width=0.49\columnwidth]{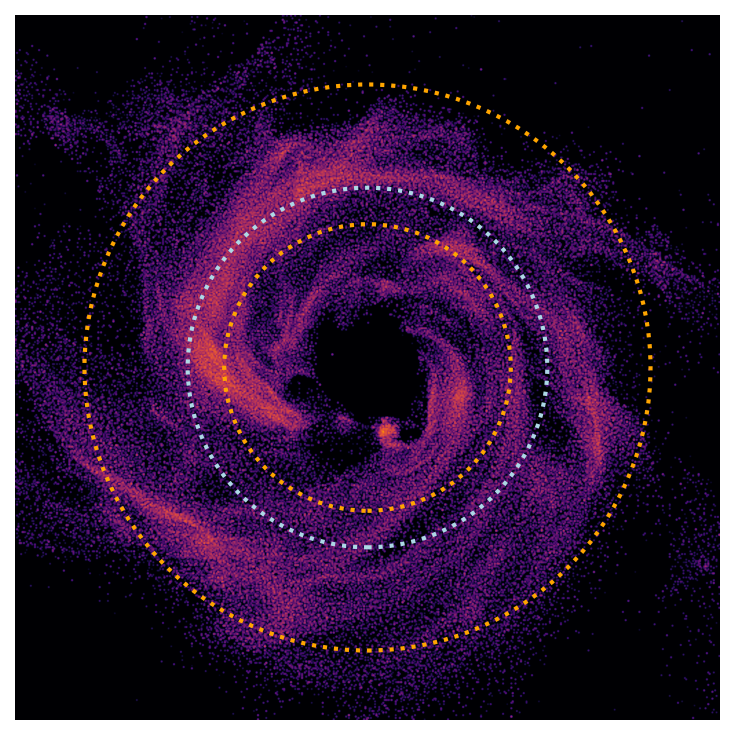}
    \includegraphics[width=0.49\columnwidth]{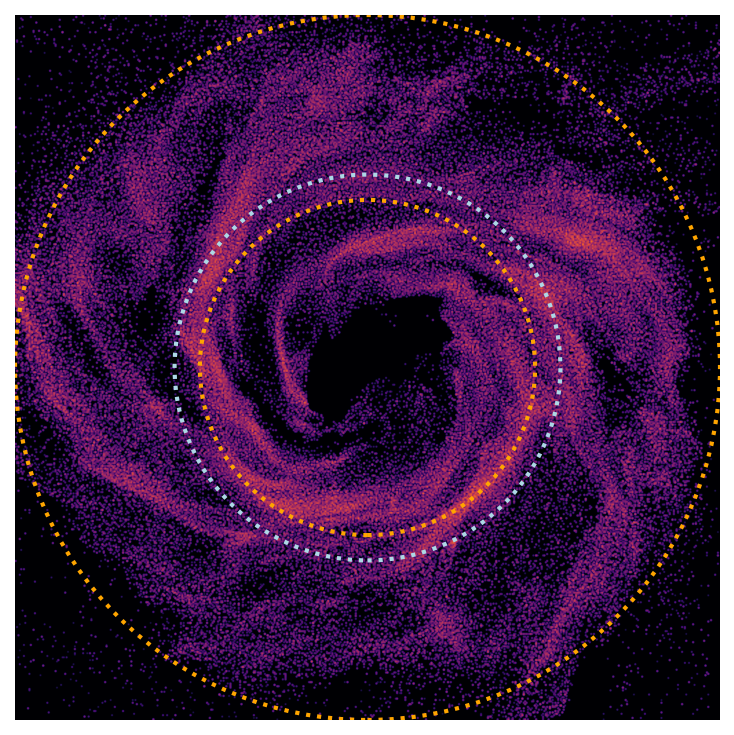}
    \caption{Mean radius ({\it blue dotted circle}) and 20th and 80th percentiles ({\it yellow dotted circles}) for the surface mass density of the gaseous rings in subhalos 167392 ({\it upper left}), 501208 ({\it upper right}), 552581 ({\it lower left}), and 595100 ({\it lower right}).}
    \label{fig:gas percentiles}
\end{figure}

However, the mean radius of the gaseous rings shows a more extended distribution compared to the stellar counterpart, with an average value of $\sim 13.87$ kpc, more than $\sim 3$ kpc larger than that of the stellar rings. This is more noticeable when comparing the mean radius with $r_{\rm hm}$, as some gaseous rings extend beyond $\sim 6$ times $r_{\rm hm}$ (Fig. \ref{fig:mean R}). The width of the rings was estimated using the 20th and 80th percentiles of the stellar particle distribution, within a range of 0.5 to $10 r_{\rm hm}$. These results show that the stellar rings are more compact than their gaseous counterparts. When comparing the 80th percentile of both components, the stellar rings have an extension of $\sim 18.21$ kpc, while the gas rings extend up to $\sim 45.25$ kpc. Figures \ref{fig:percentiles} and \ref{fig:gas percentiles} compare the mean radius, 20th, and 80th percentiles of the stellar and gaseous rings, highlighting the differences in their spatial distributions.

\subsection{Star formation rate (SFR)}

\begin{figure}
    \centering
    \includegraphics[width=\columnwidth]{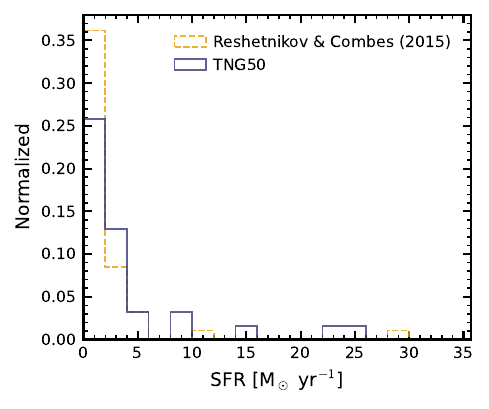}
    \caption{Distribution of the SFR. The dashed line is the SFR of observed galaxies \citep{Reshetnikov_2015}. The solid line is the instantaneous SFR of our PRG sample, estimated within $5 r_{\rm hm}$.}
    \label{fig:SFR}
\end{figure}

The SFR in PRGs is expected to be low, though higher than in typical ETGs. The bluer colour of the rings suggests the presence of ongoing star formation. Observationally, these galaxies fall in the green valley in the colour-magnitude diagram, due to the combination of the red host galaxy and the blue star-forming rings \citep{Reshetnikov_2015}. The SFRs of the TNG50 sample are mostly low (see Fig. \ref{fig:SFR}), consistent with observational data. However, a few galaxies exhibit higher SFRs, ranging from $\sim 10$ and $\sim 30 \text{\,M}_\odot \text{\,yr}^{-1}$. The highest SFR in our sample is $\sim 65.91\text{\,M}_\odot \text{\,yr}^{-1}$ (beyond the limits of Figure \ref{fig:SFR}) corresponding to subhalo 117250, which is also the most massive subhalo in the sample. 

These SFR values were measured within a radius of $5 r_{\rm hm}$ and reflect the instantaneous SFR in the gas cells\footnote{The instantaneous SFR is provided for the TNG50 simulation as a parameter of each gas cell. This star formation follows the model of \citet{2003_Springel}. Gas cells are designated as star forming when their density exceeds a specified threshold and stochastically convert to stars on a predetermined time-scale. The SFR of a gas cell is inversely proportional to the density-dependent time-scale for star formation and is proportional to the subgrid estimate of cold gas mass \citep{2019_Donnari}.}. When comparing the SFR of the subhalo with that of the ring, we find that the contribution of the ring to the overall star formation varies significantly. In many cases, the SFR in the rings constitutes a substantial fraction of the total, suggesting that the rings are the primary sites of star formation. However, also a notable number of PRGs show rings with minimal contribution to the total SFR, indicating that star formation may be occurring in other parts of the galaxy. This low star formation in some of the rings could be due to the variety of evolutionary stages present in the PRG sample from TNG50. 

\begin{figure}
    \centering
    \includegraphics[width=\columnwidth]{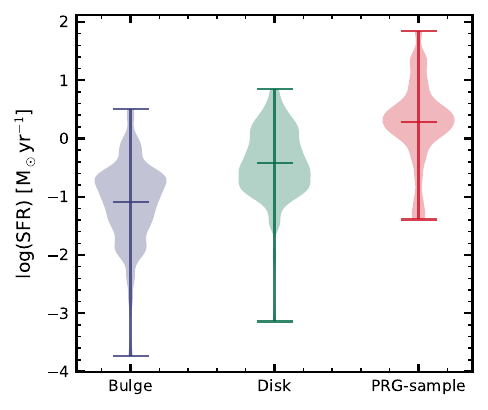}
    \caption{Comparison of the SFRs among bulge-dominated subhalos, disk-dominated subhalos, and the PRG sample. Shaded areas represent the probability density distributions, while horizontal lines indicate, from top to bottom, the maximum, mean, and minimum values.}
    \label{fig:boxplot_SFR}
\end{figure}

{To build a control sample for comparison with the SFRs of our PRG sample, we selected 100 bulge-dominated and 100 disk-dominated subhalos from the full TNG50 sample at redshift zero. This selection was based on the $\kappa$ criterion described in Section \ref{Decomposition} applied to all galaxies containing more than $10^4$ stellar particles. Figure \ref{fig:boxplot_SFR} shows violin plots of the logarithmic SFR for PRGs and the two control groups, considering only systems with non-zero SFRs. The shaded regions indicate the distribution of values for each sample, while the horizontal lines mark, from top to bottom, the maximum, mean, and minimum SFRs. For PRGs, these values are 69.42, 6.20, and 0.04 M$_\odot$ yr$^{-1}$; for the bulge-dominated control sample, 3.22, 0.25, 0.00018 M$_\odot$ yr$^{-1}$; and for the disk-dominated control sample, 7.08, 0.86, 0.00073 M$_\odot$ yr$^{-1}$. PRGs exhibit systematically higher SFRs than both control groups. This is consistent with previous observational studies, which indicate that PRGs often sustain active star formation comparable to that of star-forming disks \citep{2010_Spavone, 2012_Finkelman}. The elevated SFRs in PRGs may reflect their dynamical state. Systems with recently formed rings are expected to exhibit enhanced star formation, whereas more evolved PRGs may show reduced activity due to gas depletion or expulsion. }

\begin{table*}
\centering
\caption{(1)ID of the subhalo in TNG50, (2)Mass of the gaseous ring inside $5 r_{\rm hm}$, (3)Instantaneous SFR, (4) Instantaneous SFR of the gaseous ring, (5)Percentage of the SFR in the gaseous ring.}
\label{SFR tab}
\begin{tabular}{>{\hspace{0pt}}S[table-format=6,table-column-width=0.177\linewidth]>{\hspace{0pt}}S[table-format=2.3,table-column-width=0.202\linewidth]>{\hspace{0pt}}S[table-format=2.3,table-column-width=0.2\linewidth]>{\hspace{0pt}}S[table-format=2.3,table-column-width=0.2\linewidth]>{\hspace{0pt}}S[table-format=2.3,table-column-width=0.162\linewidth]}
\multicolumn{1}{>{\centering\hspace{0pt}}m{0.177\linewidth}}{Subhalo ID$^{(1)}$} & \multicolumn{1}{>{\centering\hspace{0pt}}m{0.202\linewidth}}{Gas Ring Mass $^{(2)}$\par{}{[}10$^{10}$ M$_\odot$]} & \multicolumn{1}{>{\centering\hspace{0pt}}m{0.2\linewidth}}{SFR total$^{(3)}$\par{}{[}M$_\odot$ yr$^{-1}$]} & \multicolumn{1}{>{\centering\hspace{0pt}}m{0.2\linewidth}}{SFR ring$^{(4)}$\par{}{[}M$_\odot$ yr$^{-1}$]} & \multicolumn{1}{>{\centering\arraybackslash\hspace{0pt}}m{0.162\linewidth}}{SFR \%$^{(5)}$} \\ 
\hline
117250 & 10.147 & 69.420 & 52.559 & 75.711 \\
143888 & 0.686 & 2.450 & 2.428 & 99.082 \\
167392 & 2.694 & 23.327 & 11.395 & 48.851 \\
264886 & 1.924 & 5.365 & 5.070 & 94.502 \\
289385 & 2.809 & 2.934 & 2.668 & 90.958 \\
313692 & 3.605 & 24.344 & 11.063 & 45.442 \\
319731 & 1.208 & 3.025 & 0.821 & 27.137 \\
324123 & 6.519 & 15.044 & 13.980 & 92.928 \\
352427 & 0.129 & 0.041 & 0.038 & 93.506 \\
383976 & 2.495 & 4.915 & 3.794 & 77.202 \\
408534 & 0.670 & 1.685 & 0.128 & 7.581 \\
411449 & 0.559 & 1.811 & 1.764 & 97.391 \\
416713 & 0.177 & 1.393 & 0.079 & 5.667 \\
428178 & 0.330 & 1.390 & 0.470 & 33.784 \\
434357 & 0.192 & 1.230 & 0.546 & 44.359 \\
483594 & 0.609 & 1.237 & 0.855 & 69.135 \\
487744 & 0.118 & 0.550 & 0.360 & 65.563 \\
501208 & 2.013 & 2.920 & 2.715 & 92.967 \\
513105 & 0.143 & 0.095 & 0.061 & 63.722 \\
514272 & 0.478 & 1.609 & 0.788 & 48.962 \\
518120 & 0.414 & 3.436 & 1.106 & 32.202 \\
532301 & 1.118 & 8.495 & 7.425 & 87.410 \\
535774 & 1.947 & 9.466 & 8.952 & 94.568 \\
536654 & 0.320 & 1.021 & 0.779 & 76.344 \\
545703 & 0.002 & 0.054 & 0.000 & 0.000 \\
552581 & 0.886 & 3.110 & 2.396 & 77.038 \\
571908 & 1.396 & 2.316 & 1.049 & 45.272 \\
585517 & 0.023 & 0.818 & 0.065 & 7.970 \\
595100 & 0.225 & 2.198 & 0.574 & 26.109 \\
596660 & 0.198 & 1.990 & 0.494 & 24.840 \\
633078 & 0.018 & 0.421 & 0.121 & 28.693 \\
649627 & 0.029 & 0.218 & 0.076 & 34.922 \\
\hline
\end{tabular}
\end{table*}

\begin{table*}
\centering
\caption{(1)ID of the subhalo in TNG50, (2)Absolute Magnitude in the $r$ band, (3) Bulge to Total {mass} fraction, (4)Mean ring radius, (5)Relation of the mean ring radius and the half mass radius, (6)Percentile 20 of the radii distribution, (7)Percentile 80 of the radii distribution, (8)Angle between the specific angular momentum of the central disk and the ring, (9)Distance from the perpendicularity of the ring.}
\label{Results}
\begin{tabular}{S[table-format=6]S[table-format=+2.3]S[table-format=1.3]S[table-format=2.3]S[table-format=1.3]S[table-format=2.3]S[table-format=2.3]S[table-format=3.3]S[table-format=2.3]}
\multicolumn{1}{c}{Subhalo ID$^{(1)}$} & \multicolumn{1}{c}{M$_{r}^{(2)}$} & \multicolumn{1}{c}{B/T$^{(3)}$} & \multicolumn{1}{c}{\begin{tabular}[c]{@{}c@{}}R$^{(4)}$\\{[}kpc]\end{tabular}} & \multicolumn{1}{c}{R/$r_{hm}^{(5)}$} & \multicolumn{1}{c}{\begin{tabular}[c]{@{}c@{}}Percentile 20$^{(6)}$\\{[}kpc]\end{tabular}} & \multicolumn{1}{c}{\begin{tabular}[c]{@{}c@{}}Percentile 80$^{(7)}$\\{[}kpc]\end{tabular}} & \multicolumn{1}{c}{\begin{tabular}[c]{@{}c@{}}Ring-Disk angle$^{(8)}$ \\{[}deg]\end{tabular}} & \multicolumn{1}{c}{\begin{tabular}[c]{@{}c@{}}Perpendicularity$^{(9)}$\\{[}deg]\end{tabular}} \\ 
\hline
117250 & -24.117 & 0.719 & 25.325 & 1.808 & 19.881 & 36.896 & 18.675 & 71.325\\
143888 & -21.609 & 1.003 & 9.175 & 2.090 & 6.748 & 19.901 & 136.646 & 46.646\\
167392 & -23.247 & 1.149 & 24.290 & 2.675 & 21.319 & 32.802 & 107.407 & 17.407\\
264886 & -22.136 & 0.946 & 8.145 & 1.409 & 5.868 & 14.888 & 106.658 & 16.658\\
289385 & -23.730 & 0.656 & 16.621 & 1.137 & 13.824 & 26.987 & 27.439 & 62.561\\
313692 & -22.988 & 0.545 & 12.188 & 1.401 & 7.837 & 26.047 & 23.789 & 66.211\\
319731 & -21.574 & 0.297 & 15.153 & 2.133 & 10.382 & 25.522 & 24.000 & 66.000\\
324123 & -23.329 & 0.748 & 15.170 & 1.359 & 11.693 & 28.473 & 107.574 & 17.574\\
352427 & -18.039 & 0.409 & 4.068 & 1.548 & 3.143 & 7.310 & 135.438 & 45.438\\
383976 & -21.671 & 0.543 & 10.331 & 1.807 & 7.242 & 23.146 & 37.555 & 52.445\\
408534 & -22.567 & 0.808 & 12.128 & 1.390 & 8.745 & 25.334 & 72.436 & 17.564\\
411449 & -22.107 & 0.427 & 7.438 & 1.252 & 6.393 & 10.488 & 102.870 & 12.870\\
416713 & -21.782 & 0.974 & 11.611 & 3.322 & 9.468 & 41.894 & 141.607 & 51.607\\
428178 & -20.357 & 0.188 & 9.737 & 3.008 & 7.527 & 15.295 & 49.335 & 40.665\\
434357 & -19.287 & 0.788 & 5.120 & 3.562 & 3.609 & 9.852 & 62.127 & 27.873\\
483594 & -21.626 & 0.368 & 11.152 & 3.152 & 8.223 & 20.729 & 40.526 & 49.474\\
487744 & -19.204 & 0.585 & 6.158 & 2.414 & 4.753 & 9.757 & 28.139 & 61.861\\
501208 & -21.952 & 0.208 & 17.011 & 2.882 & 12.962 & 26.606 & 119.271 & 29.271\\
513105 & -21.256 & 0.828 & 5.757 & 1.613 & 4.630 & 7.974 & 54.388 & 35.612\\
514272 & -21.217 & 0.469 & 10.078 & 2.553 & 8.758 & 15.219 & 132.557 & 42.557\\
518120 & -20.597 & 1.110 & 9.332 & 3.214 & 7.084 & 17.157 & 137.414 & 47.414\\
532301 & -21.475 & 0.610 & 5.020 & 1.170 & 3.742 & 8.796 & 105.852 & 15.852\\
535774 & -22.168 & 1.109 & 10.895 & 2.075 & 9.209 & 18.359 & 91.196 & 1.196\\
536654 & -21.000 & 0.656 & 6.470 & 2.523 & 5.402 & 14.193 & 96.317 & 6.317\\
545703 & -20.203 & 0.757 & 4.957 & 3.044 & 4.030 & 8.559 & 142.385 & 52.385\\
552581 & -21.540 & 0.437 & 9.274 & 2.384 & 7.450 & 13.936 & 114.152 & 24.152\\
571908 & -21.038 & 0.280 & 15.573 & 2.592 & 11.532 & 29.203 & 67.490 & 22.510\\
585517 & -20.067 & 0.482 & 4.012 & 2.703 & 3.009 & 6.674 & 43.076 & 46.924\\
595100 & -20.263 & 0.162 & 9.392 & 4.236 & 8.023 & 13.457 & 83.319 & 6.681\\
596660 & -20.172 & 0.470 & 9.015 & 3.960 & 6.855 & 14.762 & 22.729 & 67.271\\
633078 & -18.698 & 1.068 & 2.564 & 1.886 & 1.779 & 6.820 & 131.861 & 41.861\\
649627 & -19.266 & 0.644 & 2.959 & 3.102 & 2.070 & 5.731 & 79.238 & 10.762 \\
\hline
\end{tabular}
\end{table*}

\begin{table*}
\centering
\caption{Parameters estimated for the gas components (1)ID of the subhalo in TNG50, (2)Mean ring radius, (3)Relation of the mean ring radius and the half mass radius, (4)Percentile 20 of the radii distribution, (5)Percentile 80 of the radii distribution, (6)Angle between the specific angular momentum of the central disk and the ring, (7)Distance from the perpendicularity of the ring.}
\label{Gas tab}
\begin{tabular}{S[table-format=6]S[table-format=2.3]S[table-format=1.3]S[table-format=2.3]S[table-format=3.3]S[table-format=3.3]S[table-format=2.3]}
\multicolumn{1}{c}{Subhalo ID$^{(1)}$} & \multicolumn{1}{c}{\begin{tabular}[c]{@{}c@{}}R$^{(2)}$\\{[}kpc]\end{tabular}} & \multicolumn{1}{c}{R/$r_{hm}^{(3)}$} & \multicolumn{1}{c}{\begin{tabular}[c]{@{}c@{}}Percentile 20$^{(4)}$\\{[}kpc]\end{tabular}} & \multicolumn{1}{c}{\begin{tabular}[c]{@{}c@{}}Percentile 80$^{(5)}$\\{[}kpc]\end{tabular}} & \multicolumn{1}{c}{\begin{tabular}[c]{@{}c@{}}Ring-Disk angle$^{(6)}$ \\{[}deg]\end{tabular}} & \multicolumn{1}{c}{\begin{tabular}[c]{@{}c@{}}Perpendicularity$^{(7)}$\\{[}deg]\end{tabular}} \\ 
\hline
117250 & 17.137 & 1.224 & 15.633 & 40.593 & 36.301 & 53.699 \\
143888 & 7.976 & 1.817 & 6.122 & 13.435 & 140.040 & 50.040 \\
167392 & 20.603 & 2.269 & 18.210 & 31.044 & 99.547 & 9.547 \\
264886 & 15.521 & 2.685 & 10.839 & 39.612 & 98.251 & 8.251 \\
289385 & 18.928 & 1.294 & 15.148 & 34.922 & 75.318 & 14.682 \\
313692 & 21.361 & 2.456 & 17.587 & 30.349 & 43.468 & 46.532 \\
319731 & 22.145 & 3.117 & 18.580 & 28.753 & 54.914 & 35.086 \\
324123 & 22.582 & 2.023 & 17.938 & 103.737 & 101.505 & 11.505 \\
352427 & 6.116 & 2.327 & 5.297 & 23.401 & 13.597 & 76.403 \\
383976 & 18.363 & 3.212 & 13.651 & 35.424 & 63.083 & 26.917 \\
408534 & 25.673 & 2.943 & 23.220 & 40.728 & 75.610 & 14.390 \\
411449 & 8.661 & 1.458 & 7.832 & 160.217 & 6.524 & 83.476 \\
416713 & 22.766 & 6.513 & 28.488 & 61.080 & 142.888 & 52.888 \\
428178 & 14.684 & 4.536 & 12.752 & 38.408 & 58.711 & 31.289 \\
434357 & 6.635 & 4.617 & 5.734 & 10.623 & 63.373 & 26.627 \\
483594 & 15.187 & 4.292 & 12.666 & 20.940 & 50.624 & 39.376 \\
487744 & 4.559 & 1.787 & 3.719 & 24.075 & 152.850 & 62.850 \\
501208 & 19.658 & 3.331 & 14.768 & 34.979 & 116.466 & 26.466 \\
513105 & 11.234 & 3.148 & 9.582 & 21.412 & 160.003 & 70.003 \\
514272 & 18.679 & 4.732 & 18.205 & 170.190 & 118.096 & 28.096 \\
518120 & 13.532 & 4.661 & 13.320 & 119.252 & 125.765 & 35.765 \\
532301 & 6.898 & 1.607 & 5.531 & 10.925 & 30.448 & 59.552 \\
535774 & 10.145 & 1.932 & 7.878 & 16.333 & 133.017 & 43.017 \\
536654 & 9.020 & 3.517 & 7.224 & 13.677 & 102.098 & 12.098 \\
545703 & 12.225 & 7.508 & 15.186 & 28.912 & 72.119 & 17.881 \\
552581 & 9.921 & 2.550 & 7.907 & 15.616 & 105.577 & 15.577 \\
571908 & 23.231 & 3.867 & 19.022 & 43.732 & 76.548 & 13.451 \\
585517 & 8.942 & 6.025 & 10.517 & 126.354 & 49.919 & 40.081 \\
595100 & 12.850 & 5.796 & 11.164 & 23.478 & 85.376 & 4.624 \\
596660 & 12.122 & 5.325 & 11.002 & 17.541 & 32.258 & 57.742 \\
633078 & 1.355 & 0.997 & 1.060 & 45.044 & 4.096 & 85.904 \\
649627 & 5.262 & 5.517 & 8.147 & 23.230 & 79.605 & 10.395 \\
\hline
\end{tabular}
\end{table*}

\section{Discussion} \label{sec:discussion}

The analysis of our PRG sample from the TNG50 simulation provides valuable insight into the physical properties and the underlying mechanisms driving the formation and evolution of these peculiar objects. 
Our procedure successfully separated the polar structure from the host galaxy in both stellar and gaseous components. This separation allowed us to potentially study the distinct physical properties and infer the evolutionary stages of PRGs. 

Recently, \cite{2023_Smirnov} identified six PRGs from 2D synthetic images of galaxies in TNG50. This sample overlaps with ours, further validating the completeness of our PRG catalogue. 

The analysis of our PRG sample showed similarities and differences with respect to the observational data. First, our sample of 32 PRGs showed that the simulated fraction of such systems is slightly higher than the observational estimates. This discrepancy could be attributed to the broader selection criteria used in our simulations or projection effects that can complicate observational identification. Our sample was defined using dynamical criteria, which produced a wide range of relative angles. Some of these are very small with respect to the host plane, making observational distinctions challenging due to projection effects, limited spatial resolution, and low surface brightness. These factors may further explain the differences observed in the properties of the rings, such as their relative inclination with respect to their host galaxies.

Our sample also exhibited a systematic difference in the inclination between stellar and gaseous rings, a result that is consistent with the variations reported by \citet{2023_Smirnov} and \citet{2020_Khoperskov}. Rings are formed with an initial inclination relative to their host galaxy and change their inclinations as they evolve, with some systems developing counter-rotating components \citep{Lopez-Castillo+2025}. This can be seen in Table \ref{tab: kappa} with some galaxies showing a significant fraction of mass with $\epsilon_{-07}$ or figure \ref{fig:decomposition} with components marked green and purple indicating a counter-rotating component.

The morphological difference between the stellar and gaseous rings, particularly in their radial extent and thickness, could offer some insight into the physical processes at play within these structures. This variation suggests that the formation of polar rings is likely driven by multiple mechanisms rather than by a single process. For example, the more extended and diffuse nature of the gaseous rings could be indicative of continuous gas accretion. The capacity of the simulation to track these changes reinforces its value in exploring scenarios where polar rings are formed through different mechanisms, such as mergers or accretion events, providing a broader understanding of the life cycles of PRGs. Meanwhile, the discrepancy in the ring-to-host ratio is a noteworthy result. It may provide insight into the formation scenario, or alternatively reflect the impact of the physical prescriptions in the simulation, such as stellar and/or black hole feedback. Further analysis is necessary to clarify this point.

The star formation rates in our sample were consistent with the expected behaviour of observed PRGs, particularly those in early-type galaxies, where low but sustained star formation is observed. The few cases of higher star formation activity in the rings provide important insights for understanding the role of PRGs in galaxy evolution, as these systems may contribute to the rejuvenation of their host galaxies.

The use of cosmological simulations such as TNG50 offers complementary insights to observational studies by effectively capturing the distribution and morphology of PRGs over a wide range of dynamical conditions. Although some discrepancies persist between the simulated and observed PRGs, these simulations are useful tools for exploring the underlying physical processes that govern the formation and evolution of PRGs, giving some clues about related unresolved problems.

\section{Summary and Conclusions} \label{sec: conclusion}

This paper presents a catalogue of PRGs extracted from the cosmological simulation TNG50. The resolution of this simulation allows for a detailed characterisation of the objects' morphologies. In subhalos with more than $10^4$ stellar particles, we applied some filters to separate the PRG sample. First, we computed the angular momentum and inclination profiles for these subhalos, selecting those with an inclination difference exceeding $30 \deg$ (up to $150 \deg$). Then, a visual inspection of the stellar projections and line-of-sight velocity maps yielded a preliminary sample of 44 candidate PRGs.   

The resulting sample was then dynamically analysed. The one-dimensional parameters $\kappa$ \citep{Sales_2012} and the circularity \citep[$\epsilon$]{Abadi_2003} in section \ref{Decomposition} show that the majority of the subhalos in the sample are dominated by spherical components (bulge, stellar halo), which limits the extraction of information related to the ring. To overcome this, we employed a two-dimensional method. Our approach uses the cosine of the angle between the vectors of the angular momentum of the particles and the angular momentum of the central part of the subhalo ($\cos{\alpha}$) and the relation of the orbit and the circular orbit of the particles \citep{jagvaral_probabilistic_2021}. This results in a two-dimensional plane that clearly distinguishes the main dynamic structures within the subhalos. By applying a clump-finding algorithm to this plane, we were able to isolate regions of high density corresponding to the main components. Using this method, we confirmed 32 subhalos as genuine PRGs from our initial sample.

The final sample represents 1.01\% of all galaxies in the B band and 0.53\% in the r band in the TNG50 simulation, which is consistent with observational estimates \citep{1990_Whitmore, 2011_Reshetnikov, 2022_Smirnov, 2024_Mosenkov}. Furthermore, the PRGs of the sample occupy a region on the colour-magnitude diagram that lies between the red sequence and the transition to the blue cloud, a distribution that reflects the contrast between the early-type host galaxy and the blue young rings.  

The host galaxies in the PRG sample exhibit a pronounced bulge component, with a mean bulge-to-total B/T ratio of 0.64, suggesting that most host galaxies are bulge dominated, consistent with observations. However, a notable number of host galaxies are disk-dominated, indicating a diverse range of morphologies. This diversity makes our PRG sample a valuable resource for understanding galaxy evolution pathways, particularly those driven by interactions and accretion processes.

The rings in our sample exhibit average radii of 2.36 and 3.41 times the effective radius for the stellar and gaseous components, respectively. The comparatively more extended structure of the gaseous rings may suggest distinct formation or accretion histories relative to the stellar rings. Furthermore, while the rings are relatively stable and well-defined, they tend to be less massive and less perpendicular than those observed in some PRGs.

A distinctive result from our study is the difference in inclinations between the stellar and gaseous rings, with the latter often exhibiting a higher degree of perpendicularity relative to the host galaxy disk. This behaviour is consistent with the findings of \citet{2023_Smirnov} and \citet{2020_Khoperskov}, who report that polar structures adjust their inclination over time due to dynamical processes. Some rings evolve into configurations resembling counter-rotating disks, emphasising the dynamic nature of PRGs and pointing to complex formation scenarios that may involve gas accretion, satellite interactions, or minor mergers.

The ring-to-host mass ratios in TNG50 are typically lower than those reported in observed PRGs, with most rings accounting for less than 20\% of the host galaxy's stellar mass. This reduced mass, combined with a relatively low but steady star formation rate concentrated in the rings, supports the idea that PRGs can retain gas and sustain star formation at moderate levels. Consequently, the rings may play a significant role in the rejuvenation of their host galaxies.

Analysing the colour-magnitude distribution of PRGs, we found that while some fall along the red sequence, the majority occupy the green valley or the upper region of the blue cloud. This distribution suggests that PRGs represent a transitional phase between early-type, red galaxies and actively star-forming, blue galaxies, a shift driven by star formation events within the rings that significantly affect the overall colours of the host galaxies.

This study highlights the role of cosmological simulations as indispensable complements to observational studies of PRGs. Our findings suggest that PRGs are versatile and dynamic structures shaped by multiple formation mechanisms, making them powerful probes of galaxy evolution. In particular, the interplay of gas accretion and galactic interactions significantly influences their morphology and star formation histories. Our catalogue lays a robust foundation for future studies to further assess the relative contributions of different formation scenarios and to refine theoretical models of polar-ring dynamics in evolving cosmic environments.

\section*{Acknowledgements}

We thankfully acknowledge a useful and constructive report by an anonymous referee.
JGLC acknowledges the  Secretaría de Ciencia, Humanidades, Tecnología e Innovación (Secihti) for his PhD grant. 
MZA acknowledges support from SECIHTI grant number 320772.
GCG acknowledges support from UNAM-PAPIIT grant number IN110824.
The authors thankfully acknowledge computer resources, technical advice, and support provided by National Supercomputing Laboratory of Southeast Mexico, a member of the SECIHTI network of national laboratories. 
The IllustrisTNG simulations were undertaken with compute time awarded by the Gauss Centre for Supercomputing (GCS) under GCS Large-Scale Projects GCS-ILLU and GCS-DWAR on the GCS share of the supercomputer Hazel Hen at the High Performance Computing Center Stuttgart (HLRS), as well as on the machines of the Max Planck Computing and Data Facility (MPCDF) in Garching, Germany.

\section*{Data availability}
The data underlying this article will be shared on reasonable request to the corresponding author.




\bibliographystyle{mnras}
\bibliography{ref} 





\bsp	
\label{lastpage}
\end{document}